\documentclass[12pt]{article}

\usepackage{scicite}
\usepackage{amssymb,amsmath}
\usepackage{graphicx}

\usepackage{times}

\newenvironment{sciabstract}{%
\begin{quote} \bf}
{\end{quote}}

\title{Twenty-three millisecond electron spin coherence of erbium ions in a natural-abundance crystal} 

\author
{M. Le Dantec$^{1}\dagger$, M. Ran\v{c}i\'{c}$^{1}\dagger$, S. Lin$^{2}$, E. Billaud$^{1}$, V. Ranjan$^{3}$, \\
D. Flanigan$^{1}$, S. Bertaina$^{4}$, T. Chaneli\`ere$^{5}$, P. Goldner$^{6}$, A. Erb$^{7}$, \\
R. B. Liu$^{2}$, D. Est\`eve$^{1}$, D. Vion$^{1}$, E. Flurin$^{1}$, P. Bertet$^{1\ast}$\\
\\
\normalsize{$^1$Universit\'e Paris-Saclay, CEA, CNRS, SPEC, }\\
\normalsize{91191 Gif-sur-Yvette Cedex, France}\\
\normalsize{$^2$The Hong Kong Institute of Quantum Information Science and Technology,}\\
\normalsize{The Chinese University of Hong Kong, Shatin, New Territories, Hong Kong, China}\\
\normalsize{$^3$National Physical Laboratory, }\\
\normalsize{Hampton Road, Teddington, Middlesex, TW11 0LW, UK}\\
\normalsize{$^4$CNRS,  Aix-Marseille  Universit\'e,  IM2NP  (UMR  7334), }\\
\normalsize{Institut  Mat\'eriaux Micro\'electronique  et  Nanosciences de Provence,  Marseille,  France}\\
\normalsize{$^5$Univ. Grenoble Alpes, CNRS, Grenoble INP, Institut N\'eel, }\\
\normalsize{38000 Grenoble, France}\\
\normalsize{$^6$Chimie ParisTech, PSL University, CNRS, }\\
\normalsize{Institut de Recherche de Chimie Paris, 75005 Paris, France}\\
\normalsize{$^7$Walther Meissner Institut, Bayerische Akademie der Wissenschaften, }\\
\normalsize{Garching, Germany}\\
\\
\normalsize{$\dagger$ These authors contributed equally to this work.}\\
\normalsize{$^\ast$To whom correspondence should be addressed; E-mail:  patrice.bertet@cea.fr.}
}

\date{}

\begin{document} 


\baselineskip24pt


\maketitle


\begin{sciabstract}
Erbium ions doped into crystals have unique properties for quantum information processing, because of their optical transition at $\boldsymbol{1.5~\mu} \textrm{m}$ and of the large magnetic moment of their effective spin-1/2 electronic ground state. Most applications of erbium require however long electron spin coherence times, and this has so far been missing. Here, by selecting a host matrix with a low nuclear-spin density ($\textrm{CaWO}_\textrm{4}$) and by quenching the spectral diffusion due to residual paramagnetic impurities at millikelvin temperatures, we obtain an $\textrm{Er}^{\textrm{3+}}$ electron spin coherence time of $\mathbf{23}$\,ms. This is the longest electron spin coherence time measured in a material with a natural abundance of nuclear spins and on a magnetically-sensitive transition. Our results establish $\textrm{Er}^{\textrm{3+}}$:$\textrm{CaWO}_\textrm{4}$ as a leading platform for quantum networks.
\end{sciabstract}

Future quantum networks will require coherent interfaces between optical photons and other long-lived degrees of freedom or processing units. Trivalent erbium ions ($\textrm{Er}^{\textrm{3+}}$) embedded in a crystal are uniquely suited for this task. Indeed, they have an optical transition at $1.5~\mu \textrm{m}$ that is well suited for fiber-based telecommunication~\cite{saglamyurek_quantum_2015}. Their electronic ground state also forms an effective spin-1/2 with a large magnetic moment, which can couple to other quantum systems such as superconducting circuits or nuclear spins~\cite{afzelius_proposal_2013}. Thanks to these properties, they may be used for optical and microwave quantum memories~\cite{afzelius_proposal_2013,probst_microwave_2015}, as well as optical-to-microwave coherent conversion~\cite{williamson_magneto-optic_2014,fernandez-gonzalvo_coherent_2015}, enabling for instance to interface distant superconducting quantum processors. In these proposals, the erbium electron spin is magnetically coupled to a microwave resonator and is used to store and frequency-convert the quantum bits; a long coherence time is therefore essential but has not yet been reported.

Dilute paramagnetic impurities in a crystal lose phase coherence by interacting with the surrounding fluctuating magnetic moments of other paramagnetic species and nuclear spins of the host matrix. To obtain long coherence times, it is thus beneficial to use crystals that have minimal concentrations of paramagnetic impurities and a low nuclear-spin density. For instance, Hahn-echo coherence times close to 1 second were obtained with low-doped phosphorus donor spins in a chemically pure silicon crystal that was isotopically-enriched in the nuclear-spin-free $^{28}\mathrm{Si}$ isotope~\cite{tyryshkin_electron_2012}. 

However, host matrices for Rare-Earth-Ions (REIs) are often based on yttrium, such as $\mathrm{Y}_2\mathrm{SiO}_5$ and $\mathrm{YVO}_4$, and tend to have high residual REI paramagnetic impurities due to the chemical similarity amongst rare-earth elements. Moreover, Y has only one natural isotope with nuclear-spin $I=1/2$ in 100\% abundance, so it cannot be isotopically enriched to suppress nuclear magnetic noise. Therefore, it has been difficult to achieve long coherence times with magnetically-sensitive electron-spin transitions in these materials. This is especially true for Er$^{3+}$, whose electron spin transition can exhibit first order magnetic sensitivity greater than 200~GHz/T and has thus demonstrated electron spin coherence no greater than $10~\mu \mathrm{s}$ in Y based materials~\cite{probst_microwave_2015}. For this reason, most demonstrations of long spin coherence in yttrium based matrices have relied on magnetically-insensitive electron-spin transitions such as Zero-First-Order-Zeeman (ZEFOZ) or clock transitions~\cite{zhong_optically_2015,ortu_simultaneous_2018,rakonjac_long_2020}. Although such approaches have demonstrated coherence times up to $5$\,ms in Yb$^{3+}$:YSO and $2$\,s in Bi:Si~\cite{wolfowicz_atomic_2013,ranjan_multimode_2020}, the required transition frequencies are often highly constrained, leading to potential incompatibility with quantum technologies that require tuneable or broadband microwave transitions. 

Here we use a non-yttrium based crystal, namely calcium tungstate (CaWO$_4$), as the host matrix for $\mathrm{Er}^{3+}$, and demonstrate long electron spin coherence times without having to resort to ZEFOZ transitions nor isotopic purification. Indeed, most nuclei in CaWO$_4$ have no spin. Only the $^{183}$W isotope of tungsten (14\% natural abundance) contributes significantly to magnetic noise, with nuclear spin $I=1/2$ and a relatively low gyromagnetic ratio of 1.8 MHz/T. This results in one of the lowest nuclear-spin densities among natural-abundance crystals, leading to recent predictions that an electron spin coherence time of $20$\,ms should be achievable in this matrix~\cite{kanai_generalized_2021}. Calcium tungstate has a tetragonal body-centered structure with lattice constants $a=b=0.524$\,nm and $c=1.137$\,nm (as shown in Fig. 1A), in which $\mathrm{Er}^{3+}$ ions substitute $\mathrm{Ca}^{2+}$ with compensation of the additional positive charge occurring in the crystal. At low temperatures, only the two lowest energy levels are occupied. Known as a Kramers doublet, they form an effective electron spin $S=1/2$ system whose g-tensor $\mathbf{g}$ is diagonal in the crystal frame with values $ \mathrm{g}_{aa} = \mathrm{g}_{bb} = 8.38 \equiv g_{\perp}$ and $\mathrm{g}_{cc} = 1.247 \equiv \mathrm{g}_{\parallel}$ \cite{mims_spectral_1961,antipin_a._paramagnetic_1968}. In this paper we will consider only the zero-nuclear-spin isotopes of erbium, thus excluding $^{167}\mathrm{Er}$. The large magnetic moment in the $ab$ plane - up to 4 times larger than a free electron - makes it particularly interesting for coupling to superconducting circuits~\cite{haikka_proposal_2017}.

The sample used in this study is a CaWO$_4$ crystal grown from high-purity natural-abundance materials~\cite{erb_growth_2013}. EPR spectroscopy reveals that all trivalent paramagnetic REIs are present at a level of $\sim 1-100$\,ppb; in particular, $[\mathrm{Er}^{3+}]=0.7\pm0.1$\,ppb (see Sup. Mat. 1.1 and 1.5). Such concentrations are barely detectable with standard EPR spectroscopy; therefore, we use quantum-limited EPR spectroscopy with superconducting resonators and amplifiers which offer higher detection sensitivity~\cite{bienfait_reaching_2016,ranjan_electron_2020}. More importantly, we measure the spin coherence at millikelvin temperatures, where decoherence due to residual paramagnetic impurities is quenched owing to their polarization in the ground state.

Schematic descriptions of the sample and setup are shown in Figs. 1B and C. The sample is cut from a larger CaWO$_4$ crystal into a $3\times6$\,mm$^2$ rectangular slab with $0.5$\,mm thickness in the $c$-axis direction. A dc magnetic field $B_0$ is applied parallel to the sample surface in the $ab$ crystallographic plane, with a controllable amplitude and angle $\varphi$ with respect to the $a$ axis, determined using X-ray diffraction with a precision of $\pm 2^{\circ}$. Three superconducting micro-resonators with frequency $\omega_0$ are patterned in a $50$\,nm niobium thin-film deposited on top of the crystal. They allow us to independently probe the spins at frequencies between $7$ and $8$\,GHz. They consist of an inter-digitated capacitor in parallel with a few-micron-wide wire inductor ($2~\mu$m for one resonator and $5~\mu$m for the two others)~\cite{bienfait_reaching_2016}. At the single-photon level, all three resonators have an internal loss rate $\kappa_{\text{int}}$ lower than $10^6~\mathrm{s}^{-1}$ (corresponding to an internal quality factor larger than $4\times 10^4$), which is sufficient for high-sensitivity spin detection. The resonators allow for detection of $\mathrm{Er}^{3+}$ electron spins located in the vicinity of the inductor, the latter generating an oscillating magnetic field $\mathbf{B_1}$ that couples to each spin with strength $g_0 = (\mu_B/\hbar) \langle 0 | \mathbf{S} | 1 \rangle \cdot \mathbf{g} \cdot \mathbf{\delta B_1} $. Here, $\mathbf{\delta B_1}$ is the rms vacuum fluctuations of the field at the spin location. Because of the spatial variation of $\mathbf{\delta B_1}$, $g_0 $ varies with the spin location in the plane perpendicular to the wire as shown in Fig. 1C when the dc field $B_0$ is applied along the wire ($\varphi=\varphi_{\text{w}}$). The angle made by the resonator inductor with the crystalline $a$-axis is determined from the maximum of the integrated echo lineshape as a function of $\varphi$ and corresponds to $\varphi_{\mathrm{w}}=51 \pm 3 ^{\circ}$.

Each resonator is capacitively coupled with a rate $\kappa_c$ to a measurement line through which microwave pulses are applied at the resonant frequency $\omega_0$. These pulses induce Rabi nutations of the spins at a frequency $4 g_0 \beta \sqrt{\kappa_{\text{c}}} / \kappa$, where $\beta =  \sqrt{P_{\text{in}}/\hbar \omega_0}$ is the pulse amplitude in (number of photons per second)$^{1/2}$, $P_{\text{in}}$ is the input power and $\kappa = \kappa_{\text{c}} + \kappa_{\text{int}}$ is the total energy damping rate~\cite{ranjan_pulsed_2020}. The Rabi frequency is proportional to $g_0$ and thus also varies spatially so that a given pulse amplitude applies different Rabi rotations to spins at different locations. Here we measure spin coherence using the Hahn-echo pulse sequence (Fig. 1D and 2A); it is composed of two pulses of same duration $\Delta t$ and amplitudes $\beta/2$ and $\beta$ respectively, separated by a delay $\tau$, leading to the emission of a spin echo at a delay $\tau$ after the second pulse. Because the spins contributing the most to the Hahn-echo are those undergoing rotations of first $\pi/2$ then $\pi$, an echo with pulse amplitude $\beta$ probes mainly spins with a coupling constant close to $g_0 =  \pi \kappa / (4 \Delta t \beta \sqrt{\kappa_{\text{c}}})$~\cite{ranjan_pulsed_2020}. The reflected pulses, together with the spin signals, are amplified through a low-noise detection chain consisting of a Josephson Traveling Wave Parametric Amplifier (JTWPA)~\cite{macklin_nearquantum-limited_2015} followed by a High Electron Mobility Transistor (HEMT) (Fig. 1B), and demodulated at room-temperature.

An echo-detected field sweep of the erbium lineshape (see Sup. Mat. 1.3) is shown in Fig. 1D, around $B_0 = 67.2$\,mT, for several magnetic field angles $\varphi$ in the $ab$ plane. We used a pulse power such that bulk-like spins (several microns from the surface) are probed. An approximately Lorentzian lineshape is observed, with a full-width-at-half-maximum strongly dependent on $\varphi$, reaching a minimum value of $\Gamma/2\pi = 1$\,MHz for $\varphi = \varphi_0 = 31^{\circ}$ (Fig. 1E). A similar orientation-dependent linewidth was previously observed by Mims and Gillen~\cite{mims_broadening_1966}. Mims, in particular, determined that the strong $\varphi$-dependence was due to changes in $\partial \omega/\partial E_c$, the sensitivity of the transition frequency $\omega$ to an electric field $E_c$ applied along the c-axis, with $\partial \omega/\partial E_c$ vanishing at the angle $\varphi_0$~\cite{mims_electric_1965}. Following the analysis of Mims and Gillen~\cite{mims_broadening_1966}, we get quantitative agreement for a typical electric field of 32 kV/cm, probably caused by the presence of charged defects around the erbium ions and possibly related to charge compensation. The $1$\,MHz linewidth at $\varphi=\varphi_0$ is likely due to a combination of dipolar coupling to nuclear spins, other paramagnetic species in the sample, and residual Stark shifts.

To measure the longest possible spin-coherence in this system, we first cool the sample to $10$\,mK, the base temperature of the cryostat. A Hahn-echo pulse sequence is applied with sufficient microwave power to probe bulk-like spins. The spin-echo integral $A_e$ is recorded as a function of the delay $2\tau$ (Fig. 2A, the angle is $\varphi = 47^{\circ}$, close to the wire direction). A modulation of the spin-echo envelope (ESEEM) around the overall Gaussian decay is visible, and is due to the coupling to the proximal $^{183}\text{W}$ nuclear spins (see Sup. Mat. 1.7). Due to global magnetic field noise which perturbs the phase of the echo, the data are averaged in magnitude (see Sup. Mat. 1.6). A fit to the square of the magnitude $A_e^2 = \exp^{-2 (2\tau/T_2)^x}+C$~\cite{bottger_optical_2006}, with C a spurious vertical shift due to noise rectification (see Sup. Mat. 1.6), yields a coherence time $T_2 = 23.2 \pm 0.5$\,ms and $x=2.4 \pm 0.1$. This value of $T_2$ is nearly three orders of magnitude longer than previous measurements of Er$^{3+}$ electron spin coherence in $\mathrm{CaWO}_4$ at 2K~\cite{bertaina_rare-earth_2007}. We attribute this drastic improvement to both the low residual paramagnetic impurity concentration and enhanced thermal spin-polarisation at $10$\,mK, both of which greatly reduce electronic spin-spin interactions in the crystal.  Also, this is more than an order of magnitude longer than previous state-of-the-art measurements of electron spin coherence in a natural abundance material~\cite{li_hyperfine_2020}, away from a ZEFOZ transition. 

In Fig. 2A we plot simulations of the echo signal decay calculated using the cluster-correlation expansion method~\cite{witzel_quantum_2005,yang_quantum_2008} (see Sup. Mat. 2), which describes the effect of the magnetic dipole interactions between the measured $\mathrm{Er}^{3+}$ and the $^{183}\mathrm{W}$ nuclear spin bath. The similarity between simulation and experiment indicates that the measured coherence time is mostly limited by the nuclear spin bath dynamics. This suggests in particular that instantaneous diffusion (ID)~\cite{tyryshkin_electron_2012} caused by the other erbium ions is negligible, due to the low erbium concentration and to the inhomogeneous broadening (signatures of ID were observed at $\varphi \simeq \varphi_0$ where the linewidth is narrower, see Sup. Mat. 1.8). We also note that the simple formula derived in~\cite{kanai_generalized_2021} predicts a coherence time of $8$\,ms, a factor $3$ shorter than both our simulation and experiment. This discrepancy is attributed to the difference between randomly spaced (amorphous) and regularly spaced (crystalline) nuclear spin-baths (see Sup. Mat. 2.3). 

We then measure $T_2$ as a function of cryostat temperature (Fig. 2B), and observe a decrease in coherence time with increasing temperature. Since the nuclear-spin contribution is temperature-independent in the experimental range, we attribute this decrease to spectral diffusion caused by all the other paramagnetic impurities present in similar or larger concentrations than erbium (see Sup. Mat. 1.1). Indeed, this spectral diffusion is quenched at low temperatures when most paramagnetic impurities are highly polarized into their ground state~\cite{takahashi_quenching_2008,rancic_coherence_2018,li_hyperfine_2020}. We use the theoretical nuclear spin decoherence curve of Fig. 2A to extract quantitatively the paramagnetic contribution; Fig. 2B shows that the latter is suppressed by a factor $\sim 20$ by cooling the sample from $500$\,mK to $10$\,mK.  

We now turn to measurements of the longitudinal spin relaxation time $T_1$, utilizing an inversion-recovery pulse sequence. A Hahn-echo measures the longitudinal polarization at delay $T$ after application of a first pulse of the same amplitude and duration as the refocusing pulse. We use CPMG (Carr-Purcell-Meiboom-Gill sequences) to increase the signal-to-noise ratio (see Sup. Mat. 1.10). The echo integral is shown in Fig. 3A as a function of $T$, for various pulse amplitudes $\beta$. The data are well fitted by an exponential, yielding a spin relaxation time $T_1$ that is strongly dependent on $\beta$. Figure 3B shows $T_1(\beta)$ for two resonators. An approximately quadratic increase of $T_1$ with $\beta$ is observed for small $\beta$, followed by a saturation at a maximum value for larger $\beta$.

This dependence of $T_1$ on $\beta$ can be understood qualitatively by the competition between two relaxation channels: the Purcell relaxation rate $\Gamma_P = 4g_0^2 / \kappa$~\cite{bienfait_controlling_2016}, and the spin-lattice relaxation rate $\Gamma_{sl}$. Low-$\beta$ measurements probe spins with a large $g_0$, close to the inductive wire, where Purcell relaxation dominates, whereas spin-lattice relaxation becomes the limiting rate for weakly coupled spins, far from the resonator, measured with large $\beta$. This is validated by simulations that take into account the distribution of theoretical coupling constants as well as Purcell and spin-lattice relaxation~\cite{ranjan_pulsed_2020}. We obtain quantitative agreement with the data for all three resonator geometries, where the attenuation of the input line and $\Gamma_{sl}$ are the only adjustable parameters (data for two resonators are shown on Fig. 3B). The spin-lattice relaxation times measured for the three resonators (at high power) show a frequency dependence compatible with the expected $\omega^{-5}$ dependence for the direct-phonon process in a Kramers ion~\cite{abragam_electron_2012} (Fig. 3B). Due to the anisotropy of the spin-lattice coupling, the relaxation rate in Kramers ions is itself often anisotropic~\cite{abragam_electron_2012,antipin_aa_anisotropy_1981}; here, we also observe such anisotropy in the high-power relaxation time, whose dependence on $\varphi$ shown in Fig. 3C is well accounted for by the model described in ref.~\cite{antipin_aa_anisotropy_1981}.

Our observation of a near-nuclear-spin-limited coherence time of $23$\,ms in $\mathrm{Er}^{3+}$:$\mathrm{CaWO}_4$ at millikelvin temperature is meaningful for several reasons. It proves that decoherence due to paramagnetic impurities in a REI-doped crystal can be almost entirely suppressed by spin-polarising the impurities through cooling, which places $\mathrm{Er}^{3+}$:$\mathrm{CaWO}_4$ amongst some of the most coherent spin systems currently known, such as donors in isotopically purified silicon. It also confirms the predictions that CaWO$_4$ is among the best natural-abundance host crystals for long-coherence time paramagnetic defects~\cite{kanai_generalized_2021}. We note that the growth of a calcium tungstate crystal enriched in the nuclear-spin-free tungsten isotopes is also feasible, and should lead to even longer coherence times. Combined with the observation of Purcell spin relaxation, our work establishes $\mathrm{Er}^{3+}$:$\mathrm{CaWO}_4$ as a leading platform for implementing a microwave quantum memory as well as microwave-to-optical conversion. 




\section*{Acknowledgments}
We acknowledge technical support from P.~S\'enat, D. Duet, P.-F.~Orfila and S.~Delprat, and are grateful for fruitful discussions within the Quantronics group. We acknowledge IARPA and Lincoln Labs for providing the Josephson Traveling-Wave Parametric Amplifier. \textbf{Funding:} This project has received funding from the European Union's Horizon 2020 research and innovation program under Marie Sklodowska-Curie Grant Agreement No. 765267 (QuSCO) and No. 792727 (SMERC). E.F. acknowledges support from the ANR grant DARKWADOR:ANR-19-CE47-0004. We acknowledge support from the Agence Nationale de la Recherche (ANR) through the Chaire Industrielle NASNIQ  under contract ANR-17-CHIN-0001 cofunded by Atos and through the project MIRESPIN under contract ANR-19-CE47-0011, and of the Region Ile-de-France through the DIM SIRTEQ (REIMIC project). This work was supported by the ANR-Hong Kong RGC Joint Scheme (ANR-
17-CHIN-0001 and A-CUHK403/15). S.L. was supported by the Impact Postdoctoral Fellowship of CUHK. S.B. thanks the support of the CNRS research infrastructure RENARD (FR 3443). \textbf{Author contributions:}
M.L.D., M.R., E.F. and P.B. designed the experiment. A.E. grew the CaWO$_4$ crystal. The sample was provided by S.B. and was cut, polished and analysed with X-ray diffraction by P.G. S.B. performed the EPR spectroscopy at $8$\,K. M.L.D. fabricated the resonators with the help of E.B. and D.V. M.L.D. and M.R. performed the measurements with help from D.V., D.F. and P.B. M.L.D., M.R., E.F. and P.B. analysed the data. S.L. and R.B.L. performed the CCE simulations. M.L.D. performed the relaxation time simulations with the help of V.R. and P.B. M.L.D., M.R. and P.B. wrote the manuscript. T.C., P.G., S.B., D.F., D.V. and D.E. contributed to useful input to the manuscript. P.B. and E.F. supervised the project.

\section*{Supplementary materials}
Supplementary Text\\
Figs. S1 to S12\\
Tables S1 to S2\\
References \textit{(33-40)}

\begin{figure*}[tbh!]
\centering
\includegraphics[width = 16cm]{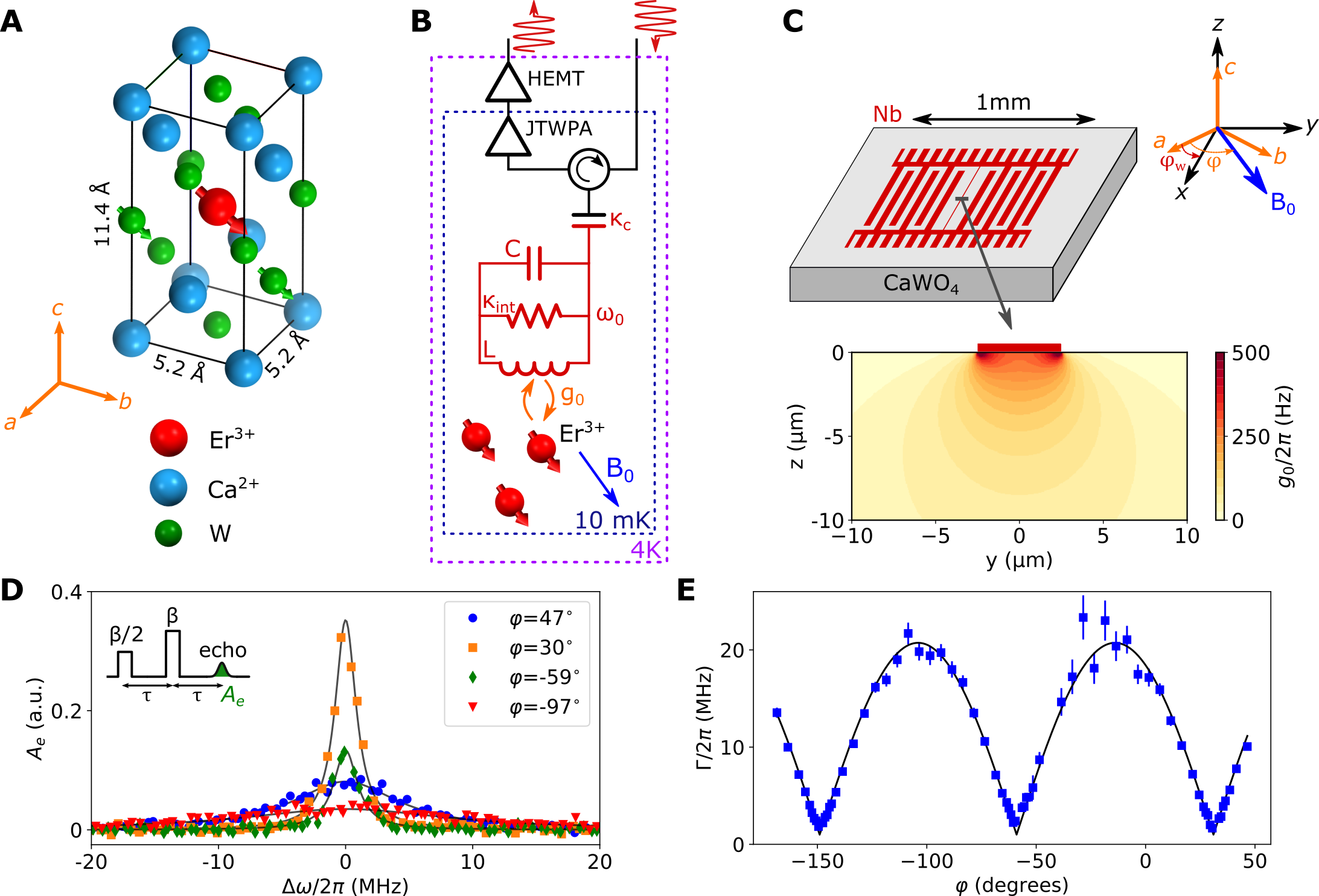}
\caption{\textbf{Schematics of the experiment and erbium spins spectroscopy.} (\textbf{A}) Unit cell of $\textrm{CaWO}_{4}$ with a central $\text{Er}^{3+}$ dopant. Oxygen atoms are removed for clarity. A fraction $0.14$ of tungsten atoms has a nuclear spin. (\textbf{B}) Experimental EPR setup. The erbium spins, subjected to a magnetic field $B_0$, are coupled with strength $g_0$ to an $LC$ resonator. The latter has an internal loss rate $\kappa_{\mathrm{int}}$ and is coupled to a measurement line with rate $\kappa_\mathrm{c}$. Microwave pulses are sent to the device, and the reflected signal containing spin-echo is routed towards a JTWPA (Josephson Traveling Wave Parametric Amplifier), followed by a HEMT (High Electron Mobility Transistor) at $4$\,K and by further amplification and demodulation at room-temperature. (\textbf{C}) Sketch of one of the three 50 nm thick niobium $LC$ resonators fabricated on top of the $\textrm{CaWO}_{4}$ sample, in the $ab$ plane. The dc magnetic field $B_0$ is applied in the $ab$ plane at
an angle $\varphi$ with respect to the $a$-axis, the resonator inductor making an angle $\varphi_{\mathrm{w}}$ with this axis. The cross-section shows the coupling $g_0$ between the resonator and erbium spins around the $5~\mu m$-wide inductance wire when $B_0$ is applied along its direction (x-axis). (\textbf{D}) Spin-echo integral $A_e$ as a function of $B_0$, around $67.2$\,mT, converted into a frequency detuning $\Delta \omega$. Full symbols are measurements for various values of $\varphi$, whereas solid lines are Lorentzian fits to the data. (\textbf{E}) Full-width-at-half-maximum linewidth $\Gamma/2\pi$ as a function of $\varphi$. The solid line is a fit following the model of~\cite{mims_broadening_1966,mims_electric_1965}, yielding a typical magnitude of inhomogeneous electric fields along the $c$-axis of $32$\,kV/cm.}
\label{fig1}
\end{figure*}

\begin{figure}[tbh!]
\centering
\includegraphics[width = 8cm]{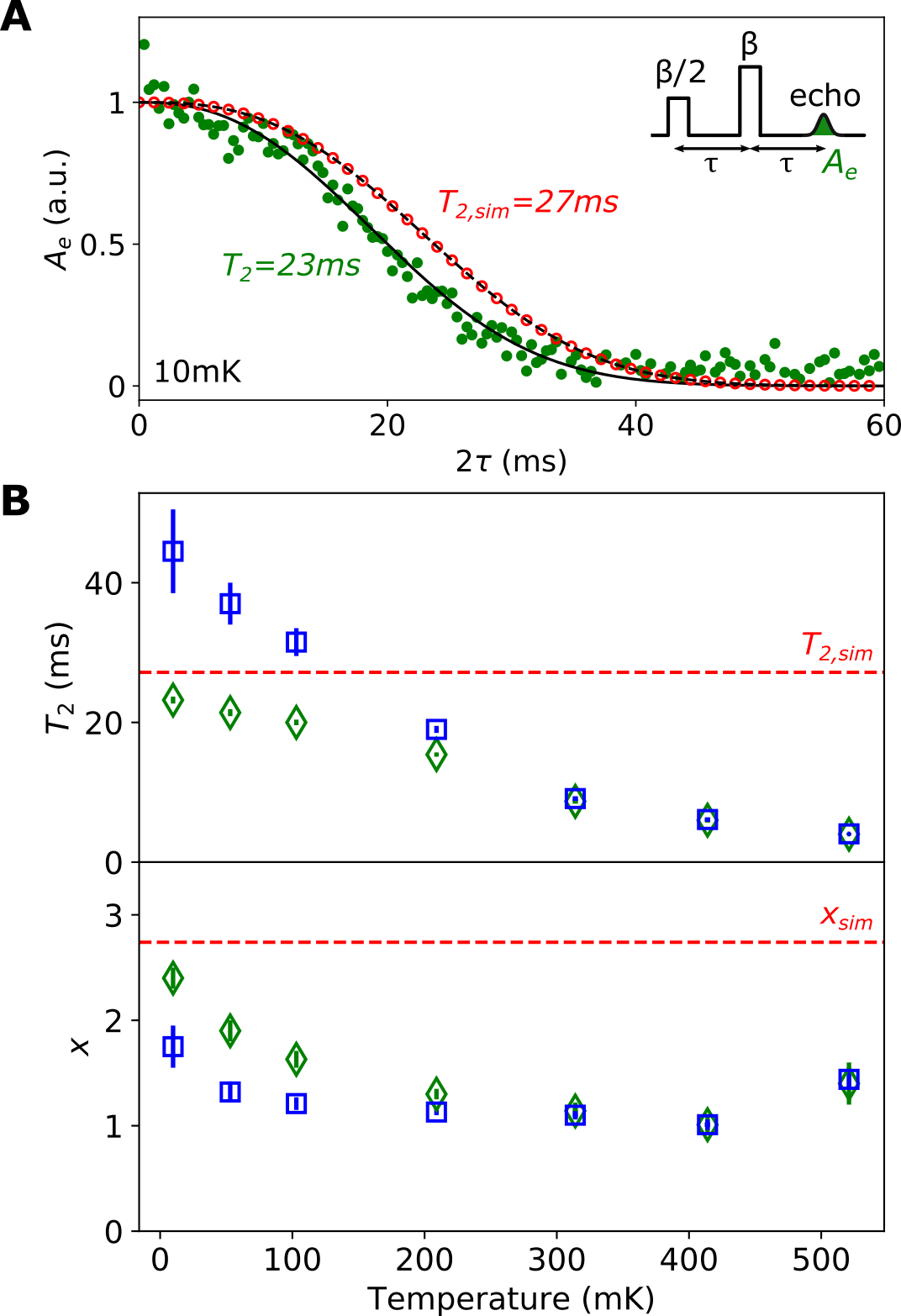}
\caption{\textbf{Er$^{3+}$ electron spin coherence time $\boldsymbol{T_2}$.} (\textbf{A}) Measured normalized Hahn-echo integral $A_e$ (green full circles) as a function of the delay $2 \tau$ between the first pulse and the echo, at $10$\,mK cryostat temperature and $\varphi = 47^{\circ}$. Each data point is magnitude averaged over $60$ measurements with a repetition time of $4$ seconds. The solid black line is a fit to $A_e^2=e^{-2(2\tau/T_2)^x}+C$ (with the offset $C$ subtracted from the data and the fit), yielding $T_2=23.2\pm0.5$ ms and $x=2.4\pm0.1$. Open red circles are the result of a cluster-correlation expansion (CCE) simulation of the nuclear spin bath for the same field orientation. The dashed black line is a fit to the simulation, yielding $T_{2,\text{sim}}=27.2$ ms and $x_{\text{sim}}=2.74$. (\textbf{B}) Measured coherence time $T_2$ and exponent $x$ (green diamonds) as a function of the cryostat temperature. The red dashed line is the result of the CCE simulation. The blue squares result from a second fit of the data as $A_e^2=e^{-2[(2\tau/T_{2,\text{sim}})^{x_{\text{sim}}}+(2\tau/T_2)^x]}+C$ in order to extract the net decoherence effect of spectral diffusion due to paramagnetic impurities.}
\label{fig2}
\end{figure}

\begin{figure}[tbh!]
\centering
\includegraphics[width = 8cm]{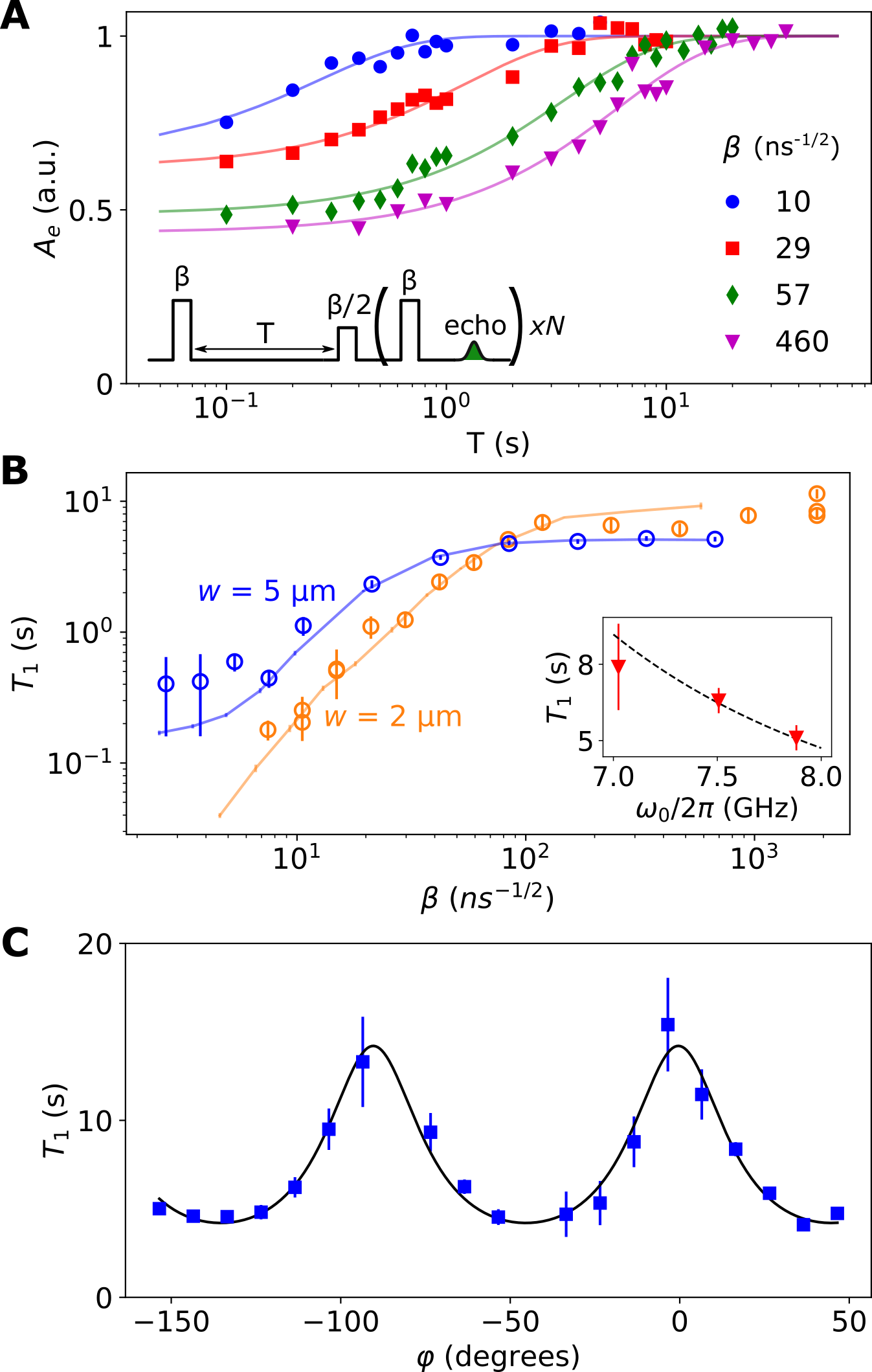}
\caption{\textbf{Spin relaxation time $\boldsymbol{T_1}$.} (\textbf{A}) The inversion recovery sequence includes a first pulse of amplitude $\beta$, followed after a delay $T$ by a Hahn-echo detection sequence also of amplitude $\beta$. The last pulse and the echo are repeated $N$ times for enhanced signal averaging (see Sup. Mat. 1.10). Solid symbols are the measured echo integral $A_e$ as a function of $T$, for various values of $\beta$ shown in the figure. Solid lines are exponential fits, yielding the spin relaxation time $T_1$. The data are measured with the $2~\mu m$-wide-inductor resonator. In (A) and (B), $\varphi$ was set to $30^{\circ}$ in order to maximize the signal. (\textbf{B}) Measured values of $T_1$ as a function of $\beta$ for a  $2~\mu m$ and $5~\mu m$-wide-inductor resonator (open circles). The solid lines result from simulations where the only adjustable parameters are the input line attenuation and the spin-lattice relaxation time (see Sup. Mat. 1.11). Inset shows the measured phonon-limited $T_1$ for all three resonators (red triangles). The black dashed line indicates that the data is compatible with a dependence of $\Gamma_{sl}$ as $\omega_0^{5}$ \cite{abragam_electron_2012}. (\textbf{C}) Measured $T_1$ (squares) at high input pulse amplitude ($\beta=700$ ns$^{-1/2}$) as a function of $\varphi$ for the $5~\mu m$-wide-inductor resonator. The solid black line is a fit with $(T_1)^{-1}=A+B\sin{(4\varphi+\varphi_1)}$, as described in \cite{antipin_aa_anisotropy_1981}, where $\varphi_1$ is found to be $92\pm3^{\circ}$.}
\label{fig3}
\end{figure}

\clearpage

\section*{\Large{Supplementary Materials}}

\renewcommand{\thefigure}{S\arabic{figure}}
\renewcommand{\thetable}{S\arabic{table}}  

\section{Experimental part}

\subsection{Standard EPR spectroscopy at $\mathbf{8}$\,K}

The identification of the impurities present in CaWO$_4$ is achieved using a conventional EPR spectrometer (Bruker EMX) operating at $9.63$\,GHz. The angular dependence is measured using an automatic goniometer with a relative resolution $<1^{\circ}$. The sample is mounted such that 60$^{\circ}$ (150$^{\circ}$) corresponds to $B_0$//a ($B_0$//c). A temperature of $8$\,K is chosen in order to optimise the signal-to-noise ratio for a range of paramagnetic impurities with varying spin relaxation rates. 

Four Kramers rare-earth-ions (Er$^{3+}$, Ce$^{3+}$, Yb$^{3+}$, Nd$^{3+}$) are identified based on known anisotropic g-tensors~\cite{mims_electric_1965}, which are simulated with open-circles in Fig.~\ref{Bruker_spectro}. The isotropic line at $B_0=160$\,mT is attributed to Fe$^{3+}$ ions~\cite{golding_epr_1978}. The relative concentration of these species with respect to erbium is estimated within an error of $10\%$: [Yb]/[Er]=3.9, [Ce]/[Er]= 54, [Nd]/[Er]=17, [Fe]/[Er]=16. At room temperature Gd$^{3+}$ ions and point defects (vacancies, Schottky type) are also observed. Note that non-Kramers ions such as Eu$^{3+}$, Ho$^{3+}$, Tb$^{3+}$ cannot be detected by low field EPR.

\begin{figure}[tbh!]
\setcounter{figure}{0}
\centering
\includegraphics[width = 8cm]{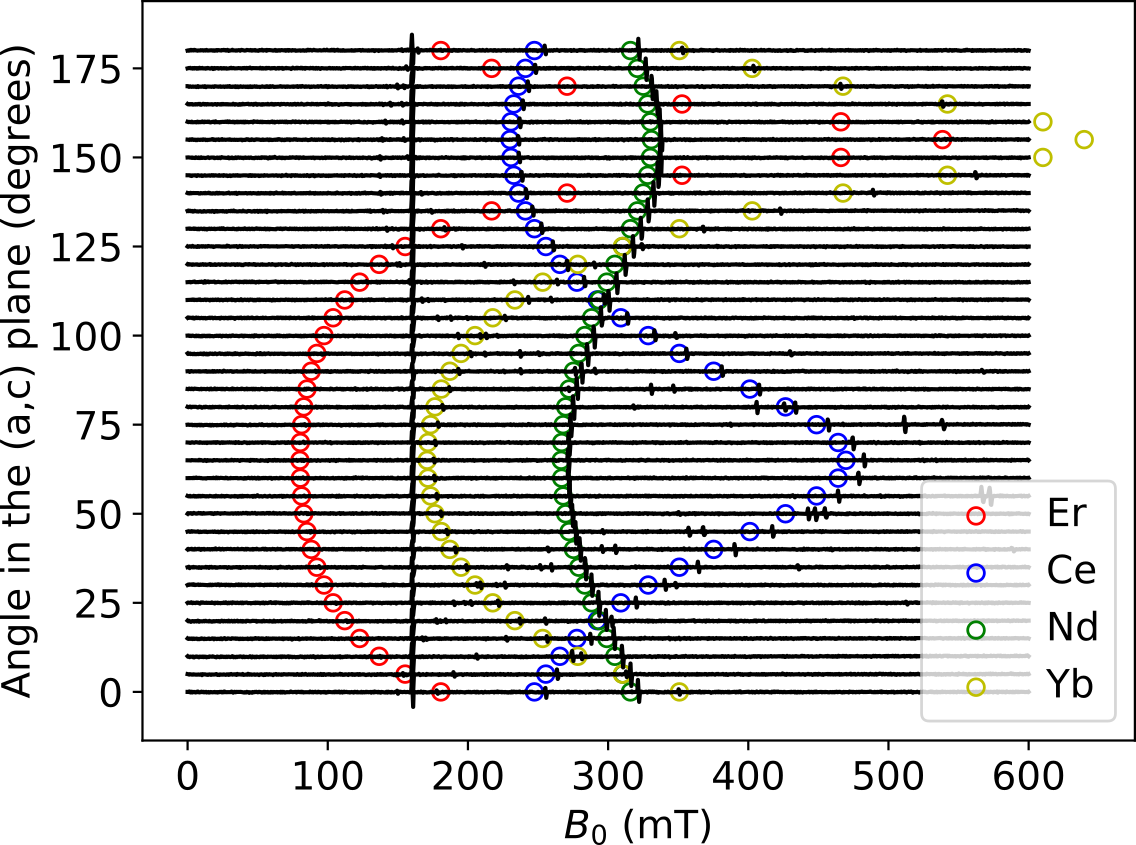}
\caption{\textbf{EPR spectroscopy of a sample taken from the same $\textrm{CaWO}_{\textrm{4}}$ boule recorded at $\boldsymbol{\omega_0/2\pi=9.63}$\,GHz and 8 K.} We identify four rare-earth ions (open circles): erbium, cerium, neodymium and ytterbium, whose g-factors in $\textrm{CaWO}_{4}$ are taken from \cite{mims_electric_1965}. The strong signal at $160$\,mT could be iron ions~\cite{golding_epr_1978}.}
\label{Bruker_spectro}
\end{figure}

\newpage

\subsection{Resonator design}

Three superconducting micro-resonators are fabricated with similar designs, in order to target slightly different frequencies. Each resonator consists in two lumped elements: an interdigitated capacitor $C$ shunted by a central inductance wire $L$. Table~\ref{geometry} shows the resonator geometric properties: wire width $w$, wire length $l$, finger width $W$ and number of pairs of interdigitated fingers $N$.

\begin{table}[tbh!]
\begin{center}
\begin{tabular}{ |c|c|c|c| } 
 \hline
 resonator design & reso 1 & reso 2 & reso 3 \\ 
 $w$ ($\mu$m) & 2 & 5 & 5 \\ 
 $l$ ($\mu$m) & 630 & 720 & 630 \\ 
 $W$ ($\mu$m)  & 10 & 50 & 10 \\
 $N$ & 8 & 6 & 8 \\
 \hline
\end{tabular}
\end{center}
\caption{\textbf{Resonator geometric properties.}}
\label{geometry}
\end{table}

\begin{figure}[tbh!]
\centering
\includegraphics[width = 5cm]{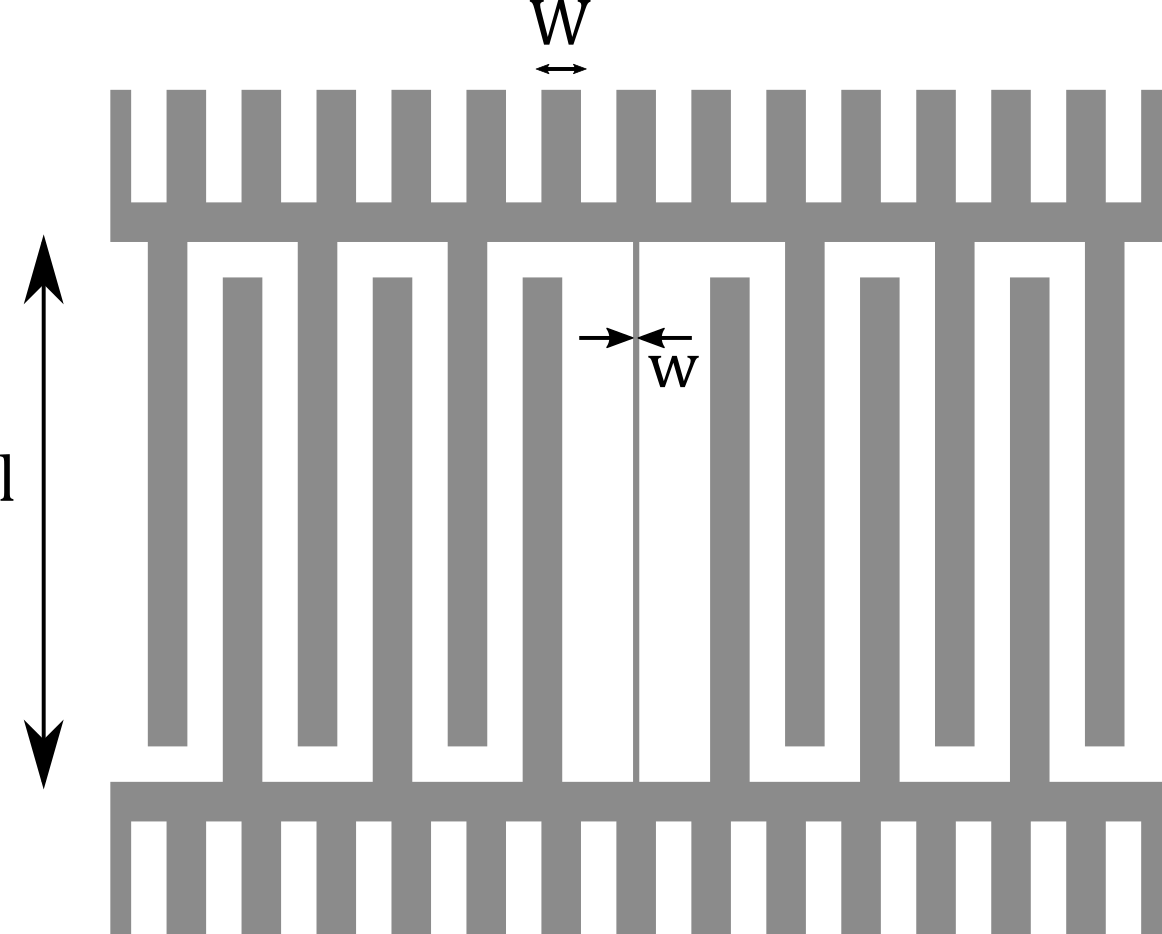}
\caption{\textbf{Example resonator design (resonator 2) with key tunable design parameters $\boldsymbol{W}$, $\boldsymbol{w}$ and $\boldsymbol{l}$ labeled. }}
\label{reso_design}
\end{figure}

Table~\ref{reso_prop} also indicates the frequency $\omega_0$, the coupling quality factor $Q_c$, the internal quality factor  $Q_i$  at single photon intra-resonator field, and total quality factor $Q_t =(1/Q_c + 1/Q_i)^{-1}$, all measured at $10$\,mK and zero magnetic field. These parameters were found to slightly vary from one experimental run to another.

\begin{table}[tbh!]
\begin{center}
\begin{tabular}{ |c|c|c|c| } 
 \hline
 resonator property & reso 1 & reso 2 & reso 3 \\ 
 $\omega_0/2\pi$ (GHz) & 7.025 & 7.508 & 7.881 \\ 
 $Q_c$ ($/1000$) & 250 & 15 & 29 \\ 
 $Q_i$ ($/1000$) & 45 & 70 & 100 \\
 $Q_t$ ($/1000$) & 38 & 12 & 22 \\
 \hline
\end{tabular}
\end{center}
\caption{\textbf{Resonator properties measured at 10 mK and zero magnetic field.}}
\label{reso_prop}
\end{table}

The resonator frequency decreases by approximately $1$\,MHz at $70$\,mT, due to kinetic inductance and an out-of-plane magnetic field component which could not be compensated for.

\subsection{EPR spectroscopy at $\mathbf{10}$\,mK and rotation pattern in the crystallographic $\mathbf{ab}$ plane}

Microwave absorption is detected using Hahn-echo sequences ($\pi/2-\tau-\pi-\tau-\text{echo}$) for a fixed magnetic field $B_0$, which is generated by two perpendicular home-made Helmholtz coils. Short delays of $\tau\sim30$-$40$\,$\mu$s are used between pulses to prevent loss of signal due to decoherence effects (refer to section 1.6 for more details). Spectra are then developed by stepping $B_0$ and repeating the Hahn-echo measurement at each step.

\begin{figure}[tbh!]
\centering
\includegraphics[width = 8cm]{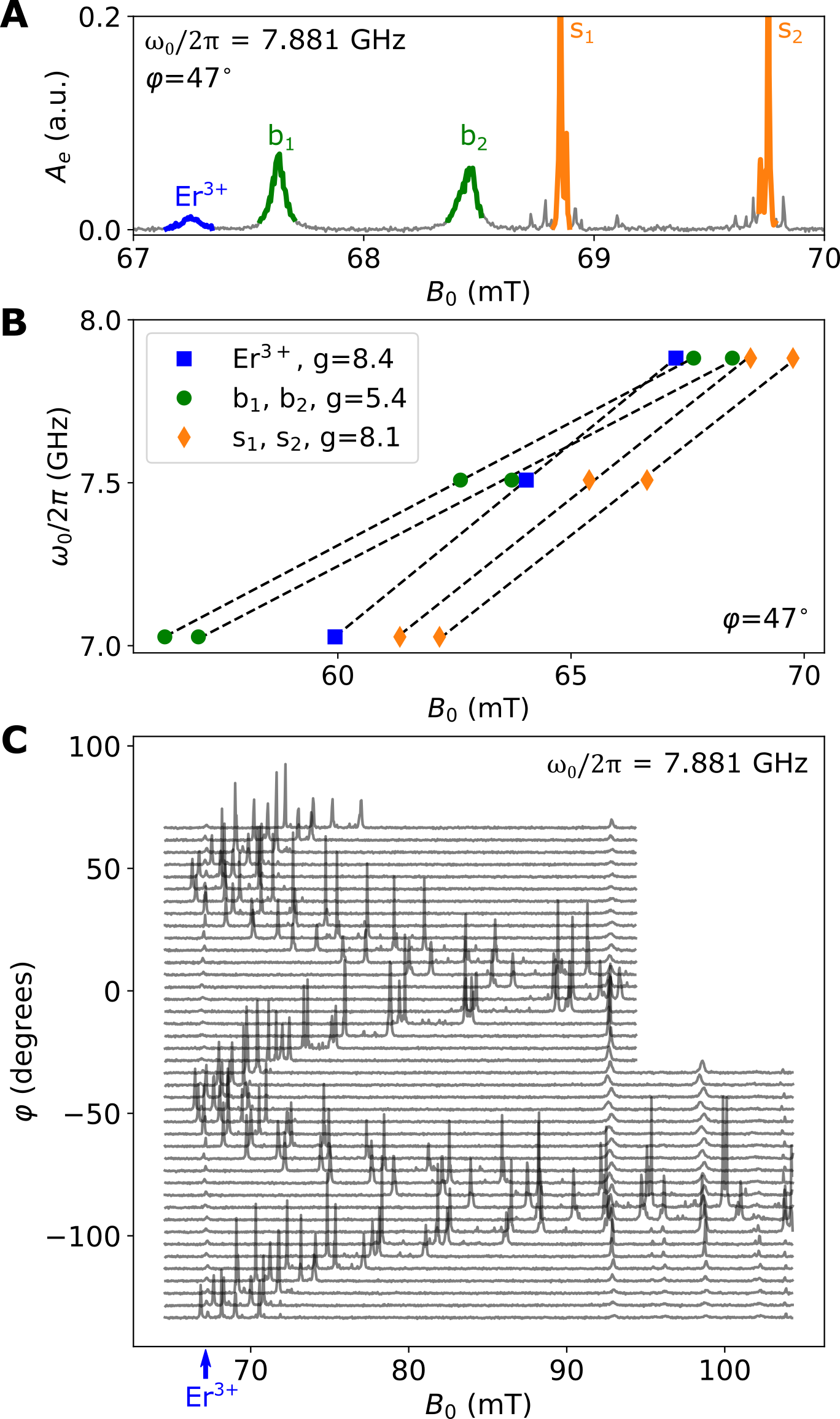}
\caption{\textbf{Hahn-echo spectroscopy at 10 mK.} (\textbf{A}) Spectrum recorded with the resonator at highest frequency $\omega_0/2\pi=7.881$\,GHz and with magnetic field angle $\varphi=47^{\circ}$. The erbium line is the peak at $67.2$\,mT. Pairs of broad peaks ($b_1$, $b_2$) and of sharp peaks ($s_1$, $s_2$) are also observed. This spectroscopy is repeated for the three resonators in order to measure the field-frequency curves in (B). (\textbf{B}) Resonance frequency as a function of the magnetic field $B_0$ for each of the five transitions detected in (A), where $\varphi=47^{\circ}$. Dashed black lines are linear fits which give the g-factor shown in the legend within a standard error of $3\%$. (\textbf{C}) Rotation pattern in the $ab$ plane with the resonator at highest frequency $\omega_0/2\pi=7.881$\,GHz. The erbium line is the smaller peak at $67.2$\,mT and its resonance frequency is independent of $\varphi$ as expected. The pairs ($b_1$, $b_2$) and ($s_1$, $s_2$) demonstrate a strong angular dependence in the $ab$ plane.}
\label{Rotation_pattern}
\end{figure}

\newpage
For the resonator at frequency $\omega_0/2\pi=7.881$\,GHz, the erbium transition is expected at $B_0=67.2$\,mT. When the magnetic field is approximately aligned with the resonator inductance wire ($\varphi \sim 50^{\circ}$), we detect other electron spin transitions in the vicinity of the erbium transition, in particular a pair of broad peaks ($b_1$, $b_2$) and sharp peaks ($s_1$, $s_2$), as shown in Fig.~\ref{Rotation_pattern}A. In order to confirm that the smallest peak at $67.2$\,mT corresponds to erbium, the spectroscopy of Fig.~\ref{Rotation_pattern}A is repeated for all three resonators. In this way, it is possible to extract the effective g-factor of each transition from the value of their frequency as a function of $B_0$. As shown in Fig.~\ref{Rotation_pattern}B, only one peak is consistent (within error) with the g-factor of erbium g$_{\perp}$=$8.38$. 

Moreover a field rotation in the $ab$ plane is performed with the resonator at frequency $\omega_0/2\pi=7.881$\,GHz in Fig.~\ref{Rotation_pattern}C and shows a strong variation of the gyromagnetic ratios of the pairs of peaks ($b_1$, $b_2$, $s_1$, $s_2$). These peaks indicate dopants in a non-tetragonal site. The slight asymmetry of the rotation pattern with respect to $\varphi=0^{\circ}$ might be caused by a weak magnetic field component along the crystal $c$-axis. Isotropic lines between $90$ and $100$\,mT are also identified, albeit without sufficient spectroscopic information to determine the impurity in question.

\begin{figure}[tbh!]
\centering
\includegraphics[width = 8cm]{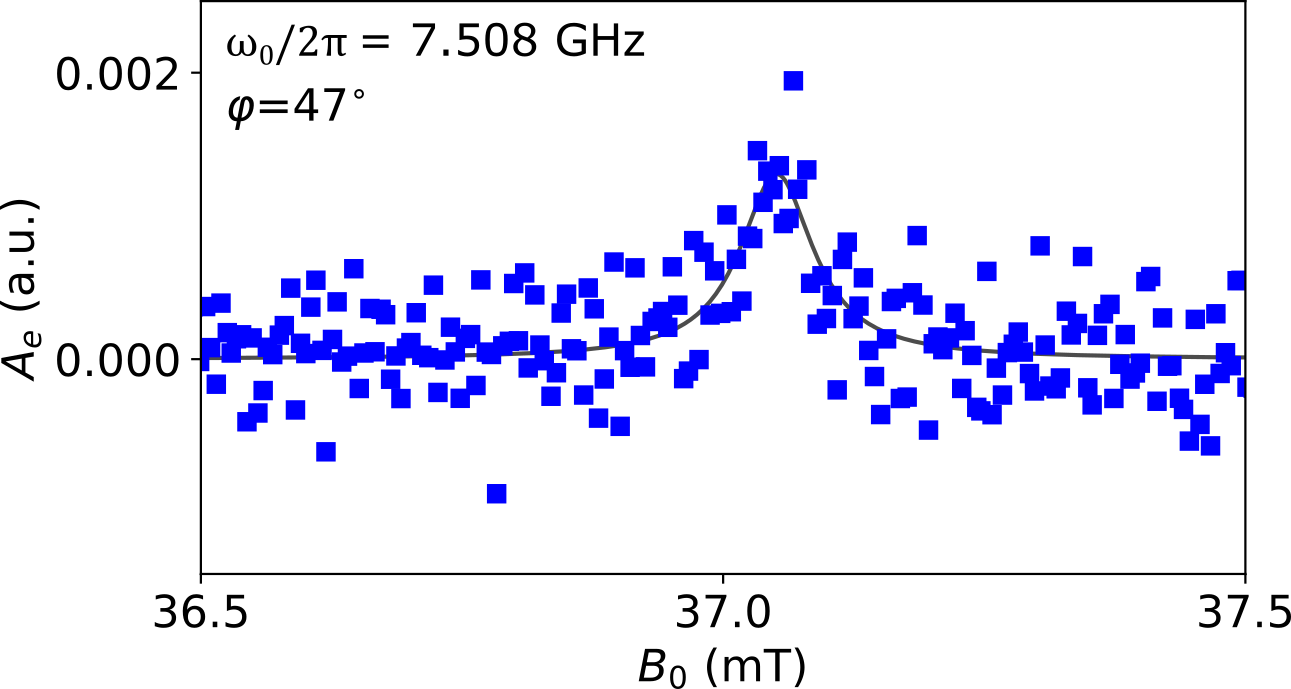}
\caption{\textbf{Hahn-echo spectroscopy over the first hyperfine transition of $^{\textrm{167}}$Er.} The Lorentzian fit (solid line) gives a full-width-at-half-maximum of $0.09\pm0.01$\,mT which corresponds to a linewidth of $\Gamma/2\pi=10\pm1$\,MHz.}
\label{hyperfine_level}
\end{figure}

Lastly, a transition at $B_0=37.05$\,mT is also observed with the resonator at $\omega_0/2\pi=7.508$\,GHz. This is consistent (within error) with the first hyperfine level of $^{167}\mathrm{Er}$. The spin-echo spectroscopy at $10$\,mK is shown in Fig.~\ref{hyperfine_level}.

\newpage

\subsection{Angular dependence of absorption linewidth}

As explained by Mims and Gillen~\cite{mims_electric_1965,mims_broadening_1966}, when the magnetic field $B_0$ is applied in the $ab$ plane, only the $c$-axis component $E_c$ of an electric field will lift the degeneracy of the g-factor g$_{\perp}$ in the $ab$ plane, such that
\begin{equation}
\delta \text{g}_{\perp}^2=2\text{g}_{\perp}\delta \text{g}_{\perp}=\alpha\sin{(2\varphi-2\varphi_0)}E_c,
\end{equation}
where $\alpha=(11\pm0.6)\times 10^{-6}~\text{(V/cm)}^{-1}$ and $\varphi_0=31\pm1^{\circ}$ for $\textrm{Er}^{3+}$:$\textrm{CaWO}_4$.

The shift in resonance frequency is then
\begin{equation}
\delta \omega = \delta \text{g}_{\perp}\frac{\mu_B}{\hbar}B_0,
\end{equation}
where $\mu_B$ is the Bohr magneton and the sensitivity of the spin-transition frequency to electric fields is
\begin{equation}
\frac{\partial \omega}{\partial E_c}=\frac{\alpha\sin{(2\varphi-2\varphi_0)}}{2\text{g}_{\perp}}\frac{\mu_B}{\hbar}B_0.
\end{equation}

Now, if each of the erbium ion sees a random electric field of the order of $\Delta E_c$ along the $c$-axis, possibly due to charge defects in the CaWO$_4$ crystal as assumed in \cite{mims_broadening_1966}, the full-width-at-half-maximum of the transition linewidth of the spin-ensemble broadens such that
\begin{equation}
\Gamma \sim \Gamma_{\text{min}}+|\frac{\partial \omega}{\partial E_c}|\Delta E_c.
\end{equation}

Fitting this formula to the data of Fig.~\ref{fig1}E leads to $\Delta E_c=32.0\pm0.6~\text{kV/cm}$. This value is approximately three times smaller than measured by Mims and Gillen in their ppm doped crystal~\cite{mims_broadening_1966}, a difference that could be attributed to our lower doping concentration, and hence reduced charge defect density. Moreover, the fit also confirms that the spin transition frequency becomes insensitive to $E_c$ at $\varphi=\varphi_0=31^{\circ}$.

\subsection{Estimation of the erbium concentration}

Determining the density of $\text{Er}^{3+}$ ions in the crystal is achieved by measuring the coupling between the erbium spin-ensemble and the microwave resonator. To do so, a complex microwave reflection measurement $r(\omega)$ was recorded on a Vector Network Analyser (VNA). As explained in ref.~\cite{diniz_strongly_2011}, when the microwave resonator is coupled to a spin ensemble of Lorentzian lineshape with coupling $g_{\text{ens}}$, the reflection coefficient can be expressed as
\begin{equation}
r(\omega) = \frac{i\kappa_c}{(\omega-\omega_0)+i\frac{\kappa_c+\kappa_{\text{int}}}{2}-\frac{g_{\text{ens}}^2}{(\omega-\omega_s)+i\frac{\Gamma}{2}}}-1
,
\end{equation}
where $\omega_0$, $\kappa_c$ and $\kappa_{\text{int}}$ are the frequency, coupling rate and loss rate of the resonator and $\omega_s$ and $\Gamma$ are the spin transition frequency and inhomogeneous linewidth, respectively.

In particular, the spin ensemble broadens the resonance linewidth according to 
\begin{equation}
\tilde{\kappa}_{\text{int}}=\kappa_{\text{int}}+\frac{g_{\text{ens}}^2\Gamma}{(\omega-\omega_s)^2+(\Gamma/2)^2}.
\end{equation}

\begin{figure}[tbh!]
\centering
\includegraphics[width = 8cm]{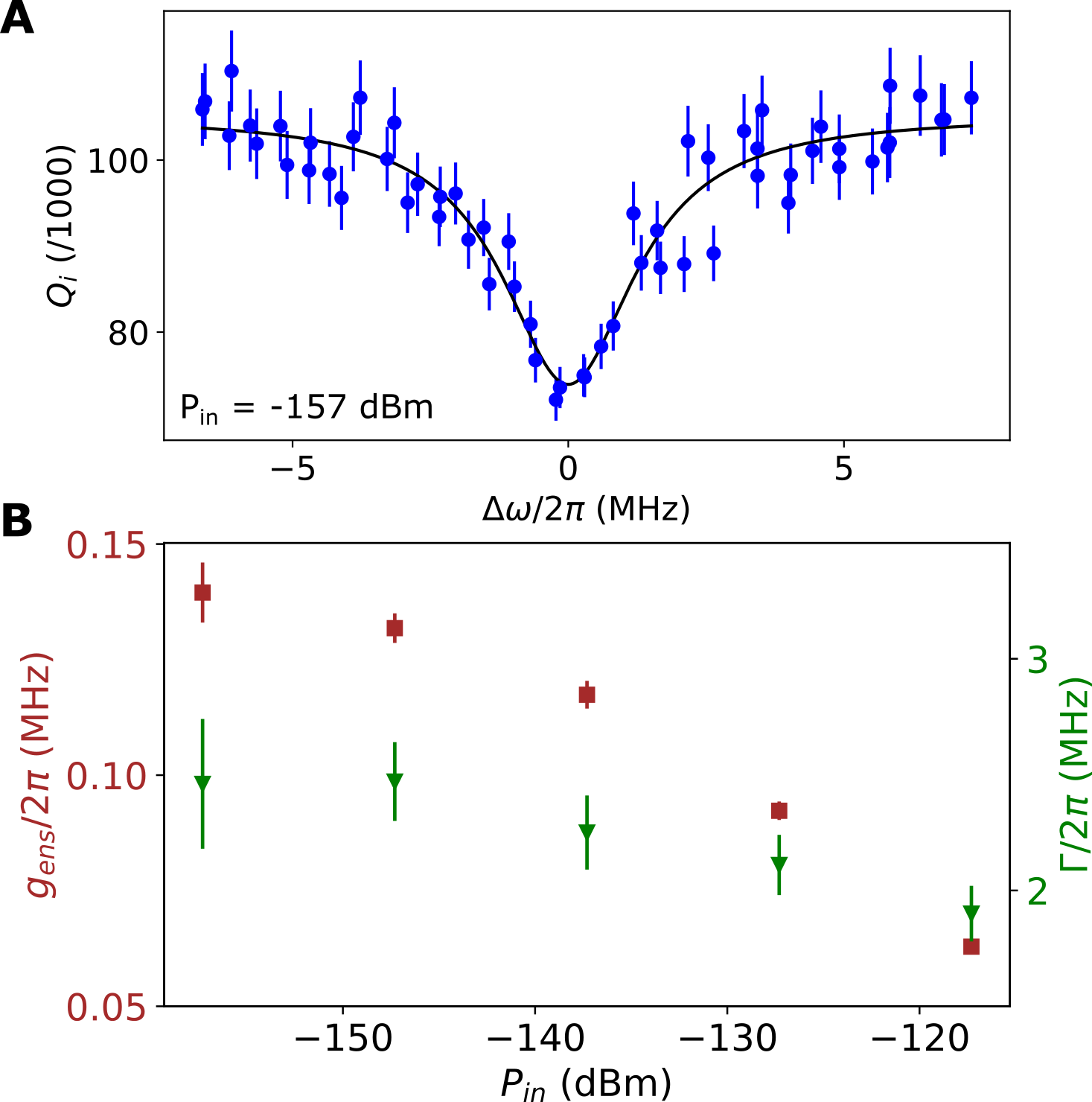}
\caption{\textbf{Continuous-wave spectroscopy at 10 mK.} (\textbf{A}) Internal quality factor $Q_i$ as a function of $B_0$, around $67.2$\,mT, and converted into a frequency detuning $\Delta \omega$, measured with input power $P_{\text{in}}$ at the sample of $-157$\,dBm. $\varphi$ is set to $30^{\circ}$. Solid black line is a fit to the data. (\textbf{B}) Fitted ensemble coupling $g_{\text{ens}}$ and linewidth $\Gamma$ as functions of $P_{\text{in}}$. As the input power decreases, the ensemble coupling $g_{\text{ens}}$ saturates at $g_{\text{ens}}/2\pi=140\pm6$\,kHz.}
\label{fig_g_ensemble}
\end{figure}

The measured internal quality factor is shown in Fig.~\ref{fig_g_ensemble} and the fit with $Q_i=\omega_0/\tilde{\kappa}_{\text{int}}$~\cite{ranjan_probing_2013} yields an ensemble coupling at low input power of $g_{\text{ens}}/2\pi=140\pm6$\,kHz.

The concentration $\rho$ of the zero nuclear-spin isotopes of erbium is then estimated by considering how $g_{\text{ens}}$ depends on the spatial variation of vacuum magnetic field fluctuations under the resonator~\cite{kubo_strong_2010}
\begin{equation}
\begin{gathered}
g_{\text{ens}}=\\
\frac{\mu_B}{2\hbar} \sqrt{\rho\int_V\mathbf{dr}(g_{\parallel}\delta B_{1z}(\mathbf{r}))^2+(g_{\perp}\cos{(\Delta \varphi)}\delta B_{1y}(\mathbf{r}))^2}
\end{gathered}
\end{equation}
where $\Delta \varphi=\varphi-\varphi_{\text{w}}=21^{\circ}$ is the angle between $\mathbf{B_0}$ and the resonator wire axis.

From this equation, we estimate $\rho=(0.7\pm0.1)\times 10^{13}$\,cm$^{-3}$. This corresponds to a total trivalent erbium concentration (including all isotopes) of $[\text{Er}^{3+}]=\frac{\rho}{0.77}=0.7\pm0.1$\,ppb.

\subsection{Homogeneous phase noise and $\mathbf{T_2}$ measurements}

We measure spin echoes using homodyne detection and obtain two field quadratures whose constant offsets are subtracted, yielding $I(t)$ and $Q(t)$. The echo has a phase $\theta$ in the $IQ$ plane which depends ideally only on the phase of the driving pulses. In order to have the best signal-to-noise ratio, the echo amplitude is often computed using an average of $N$ individual echo traces as follows
\begin{equation}
A_{e,\text{phase}}=\operatorname{Re}{(e^{-i\theta}\frac{1}{N} \sum_{n=1}^N{\{\int_t[I_n(t)+iQ_n(t)]dt\}}}).
\end{equation}

However, when using a Hahn-echo sequence ($\pi/2-\tau-\pi-\tau-$echo), we observe in our measurements that the echo phase $\theta$ varies from trace to trace when the delay $\tau$ becomes larger than about $1$\,ms. This we attribute to a time-varying homogeneous perturbation over the spin ensemble. Although we could not determine the origin of this perturbation, we suspect global magnetic field variations at kHz frequencies that can modulate the ensemble transition frequency and hence cause echo refocusing at varied angles $\theta$.

Therefore, we compute the amplitude of the echo in magnitude, so that it is not sensitive to the echo phase. We choose to compute this magnitude as  
\begin{equation}
A_{e,\text{mag}}=\sqrt{\frac{1}{N} \sum_{n=1}^N{\{[\int_t{I_n(t)}dt]^2+[\int_t{Q_n(t)}dt]^2\}}}.
\end{equation}

\begin{figure}[tbh!]
\centering
\includegraphics[width = 8cm]{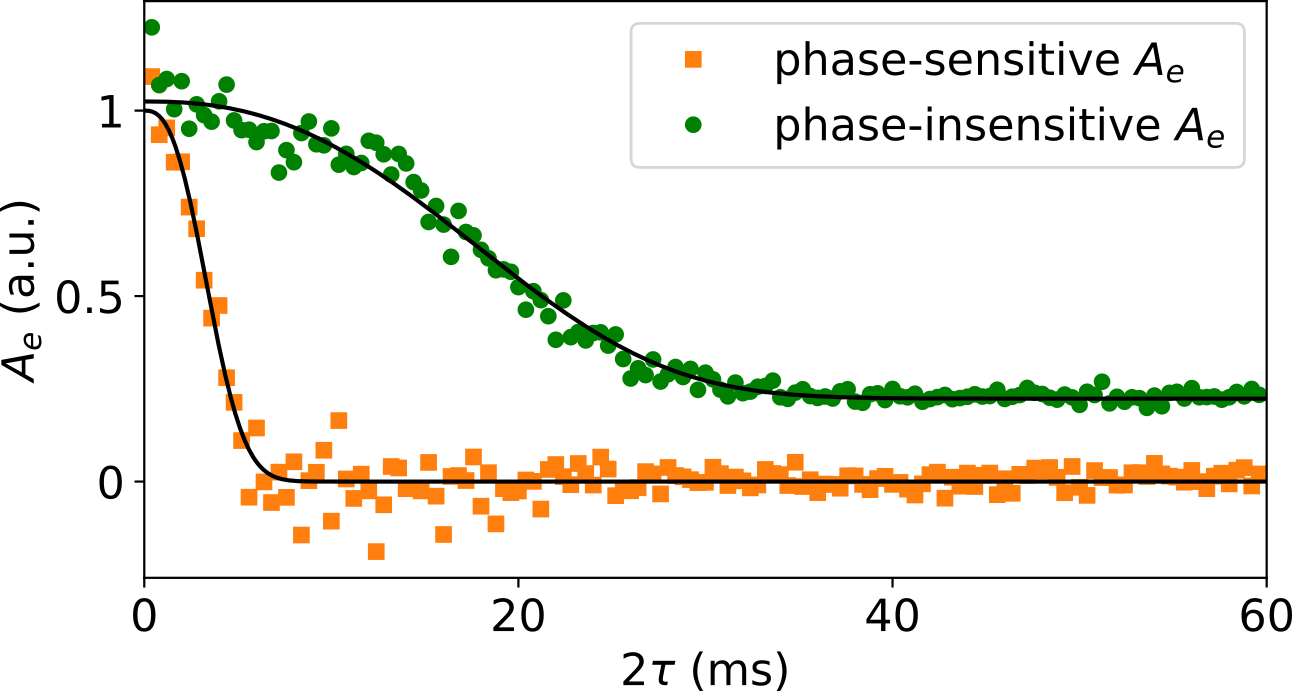}
\caption{\textbf{Electron spin coherence time measurement at 10 mK and $\boldsymbol{\varphi=47^{\circ}}$.} The data are averaged either in a phase-sensitive (squares) or insensitive (circles) manner. Solid black lines are fits, the phase-sensitive data being fitted with $A_e=e^{-(2\tau/T_{2,\text{phase}})^{x_{\text{phase}}}}$ and the phase-insensitive data with $A_e=\sqrt{e^{-2(2\tau/T_{2,\text{mag}})^{x_{\text{mag}}}}+C}$. The offset $C$ is a spurious vertical shift corresponding to the variance of the noise and which occurs when averaging the data in magnitude. The fits yield $T_{2,\text{phase}}=4.0\pm0.1$\,ms, $x_{\text{phase}}=2.6\pm0.2$, $T_{2,\text{mag}}=23.2\pm0.5$\,ms and $x_{\text{mag}}=2.4\pm0.1$.}
\label{figT2_Q_vs_ampl}
\end{figure}

Fig.~\ref{figT2_Q_vs_ampl} shows the decay of the echo amplitude with the two mentioned averaging methods. When the spin-echo integral $A_e$ is averaged in a phase-sensitive manner, the fitted coherence time $T_{2,\text{phase}}$ is $4.0\pm0.2$\,ms, whereas phase-insensitive averaging yields the correct coherence time $T_{2,\text{mag}}=23.2\pm0.5$\,ms. 

It is well known that spin-coherence can be extended using Dynamical-Decoupling (DD) techniques such as the Carr-Purcell-Meiboom-Gill (CPMG) sequence~\cite{ma_uncovering_2014}. However, DD sequences requires phase-cycled averaging to suppress spurious stimulated echoes which, in-turn, requires predictable and stable echo signal phase. Therefore, DD measurements were not attempted here due to the observed phase instability in our measurements.

With the exception of coherence time measurements, all data in this article are taken with phase-sensitive averaging, for sufficiently short delays $\tau$ yielding negligible fluctuations of the echo phase $\theta$.

\subsection{Electron spin-echo envelope modulation (ESEEM)}

\begin{figure}[tbh!]
\centering
\includegraphics[width = 8cm]{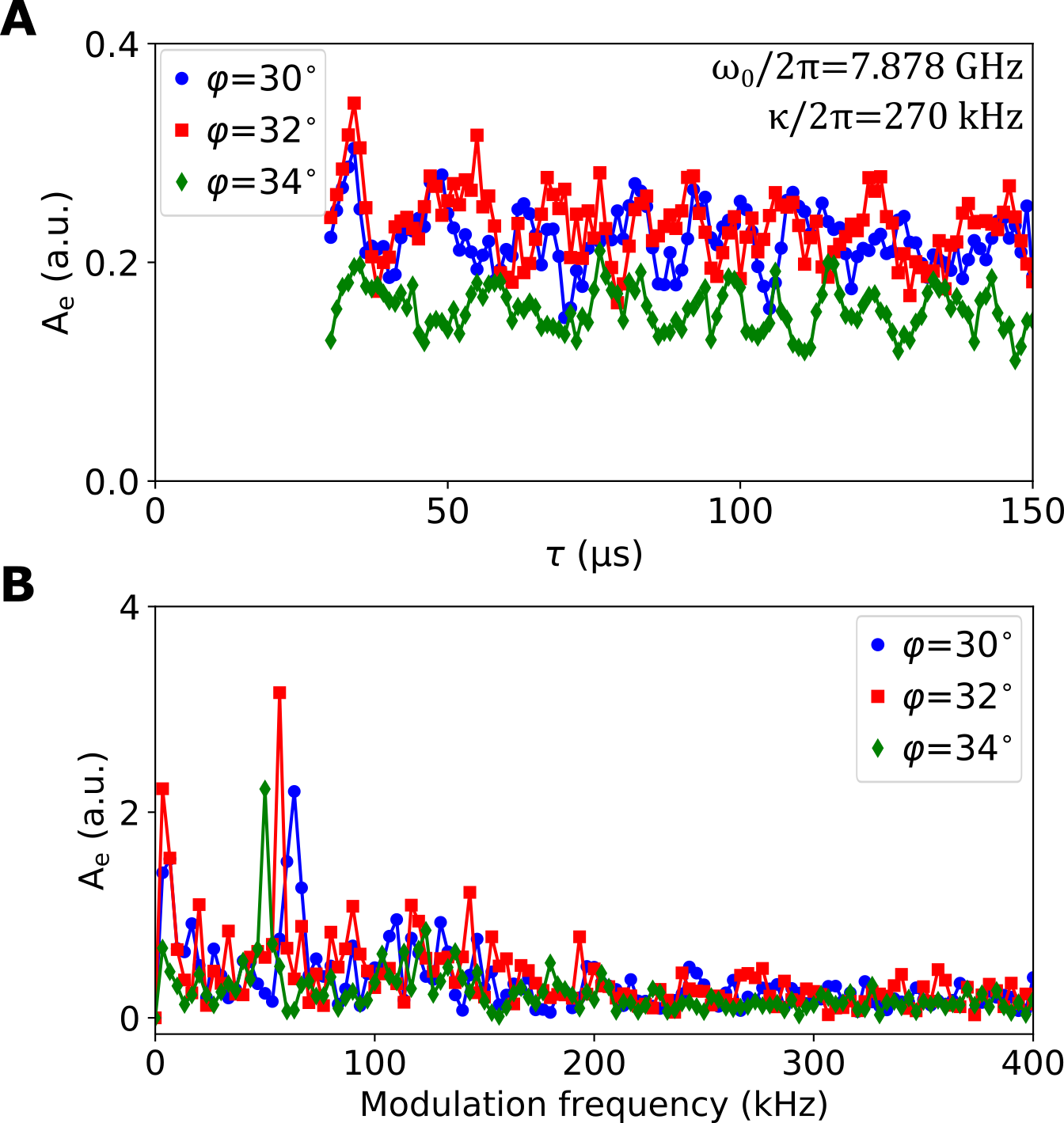}
\caption{\textbf{Electron spin echo envelope modulation measured with resonator 3.} (\textbf{A}) Echo amplitude averaged in quadrature as a function of short inter-pulse delay $\tau$, measured at $10$\,mK and with different field orientations $\varphi$. The resonance has a frequency $\omega_0/2\pi=7.878$\,GHz and a linewidth $\kappa/2\pi=270$\,kHz. $4~\mu s$-long square pulses are used such that the pulse bandwidth is $\Delta\omega_{\textrm{pulse}}/2\pi \approx 250$\,kHz. (\textbf{B}) Fast Fourier transform of the data of subplot (A).}
\label{figeseem_reso3}
\end{figure}

\begin{figure}[tbh!]
\centering
\includegraphics[width = 8cm]{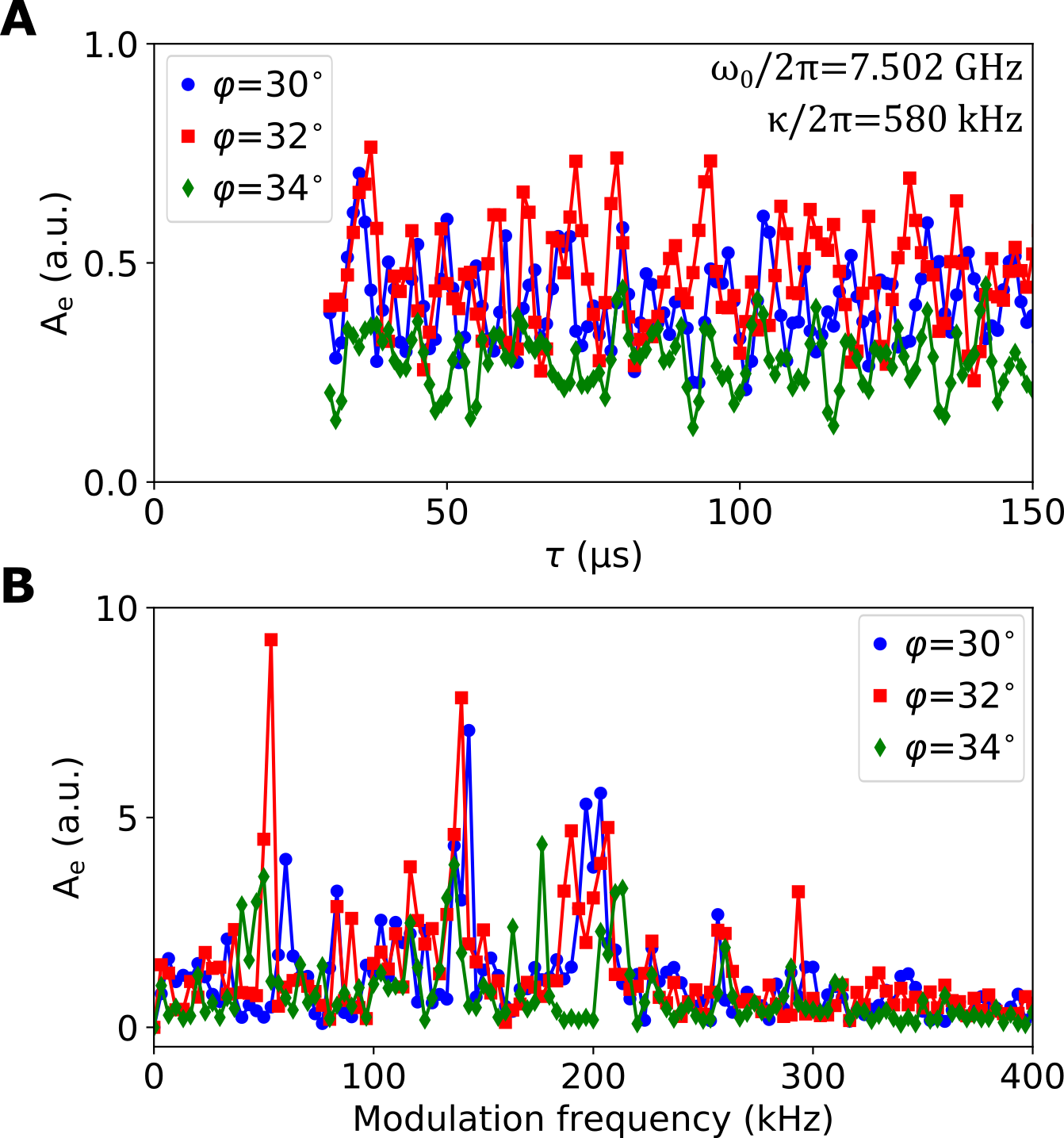}
\caption{\textbf{Electron spin echo envelope modulation measured with resonator 2.} (\textbf{A}) Echo amplitude averaged in quadrature as a function of short inter-pulse delay $\tau$, measured at $10$\,mK and with different field orientations $\varphi$. The resonance has a frequency $\omega_0/2\pi=7.502$\,GHz and a linewidth $\kappa/2\pi=580$\,kHz. $1~\mu s$-long square pulses are used such that the pulse bandwidth is $\Delta\omega_{\textrm{pulse}}/2\pi \approx 1$\,MHz. (\textbf{B}) Fast Fourier transform of the data of subplot (A).}
\label{figeseem_reso2}
\end{figure}

In Fig.~\ref{figT2_Q_vs_ampl}, the noise of the data averaged in magnitude seems to decay with the delay $2\tau$. Complementary measurements record the echo amplitude at short inter-pulse delay $\tau$, with a sampling time of $\Delta \tau=1~\mu$s (Fig.~\ref{figeseem_reso3} and Fig.~\ref{figeseem_reso2}). The time interval for these measurements is $300~\mu$s and the echo amplitude averaged either in quadrature or magnitude does not decay on this scale as $T_{2,\mathrm{phase/mag}}$ is much larger. The time traces are taken for several field orientations $\varphi$ and display a strong modulation which is evidenced by taking the fast Fourier transform of each trace. This phenomenon originates from ESEEM where the Hahn-echo decay is modulated by the coupling between the erbium ion and its neighboring tungsten atoms of $^{183}\mathrm{W}$ \cite{probst_hyperfine_2020}. As shown on Fig.~\ref{figeseem_reso3}B and Fig.~\ref{figeseem_reso2}B, the modulation frequencies are angular dependent due to the anisotropy of the dipolar coupling between erbium and tungsten. 

Moreover, the data presented in Fig.~\ref{figeseem_reso3} and Fig.~\ref{figeseem_reso2} are taken with two resonators which have slightly different frequencies and more importantly different resonance linewidths. The resonance linewidth $\kappa$ combined with the excitation pulse bandwidth $\Delta \omega_{\textrm{pulse}}$ filters out high frequencies of the theoretical ESEEM. In Fig.~\ref{figeseem_reso3}, $\kappa/2\pi=270$\,kHz and $\Delta\omega_{\textrm{pulse}}/2\pi \approx 250$\,kHz so modulation frequencies larger than $125$\,kHz are filtered out. In Fig.~\ref{figeseem_reso2}, $\kappa/2\pi=580$\,kHz and $\Delta\omega_{\textrm{pulse}}/2\pi \approx 1$\,MHz so modulation frequencies larger than $290$\,kHz are filtered out.

\newpage

\subsection{$\mathbf{T_2}$ measurements and instantaneous diffusion}

Spin coherence is determined by several electric and magnetic interactions which contribute to the decay of the echo amplitude according to
\begin{equation}
A_e(2\tau)=e^{-\sum_i{(\frac{2\tau}{T_{2,i}})^{x_i}}}
\end{equation}
where $T_{2,i}$ is the coherence time and $x_{i}$ the exponent due to interaction mechanism $i$.

In this work, we consider three dominant magnetic interactions: spectral diffusion (SD) due to the nuclear spin bath, SD due to other paramagnetic impurities and instantaneous diffusion (ID). SD has already been mentioned in the main text and comes from the magnetic interaction of the erbium ions with all other electron and nuclear spins in the bath. ID is also a common decoherence mechanism and comes from the magnetic interaction between all erbium spins which are resonant with the microwave excitation pulses. 

In this section we estimate at $\varphi=47^{\circ}$ the decoherence due to ID and compare it with the measured coherence time presented in Fig.~\ref{fig2}.

ID is expected to give an exponential decay $A_{e,\text{ID}}(2\tau)=e^{-2\tau/T_{2,\text{ID}}}$ \cite{tyryshkin_electron_2012,schweiger_principles_2001} with
\begin{equation}
\frac{1}{T_{2,\text{ID}}}=\frac{5}{2} \frac{\mu_0}{4\pi} \frac{(g \mu_B)^2}{\hbar} \rho \frac{\Delta \omega}{\Gamma} \sin^2{\frac{\theta_2}{2}},
\label{T2_ID}
\end{equation}
where $\mu_B$ is the Bohr magneton, $\rho=0.77[\text{Er}^{3+}]$ is the zero nuclear-spin isotopes erbium concentration, $\Delta \omega$ is the excitation bandwidth, $\Gamma$ the spin inhomogeneous linewidth, and $\theta_2$ the refocusing angle. 

For the measurement of Fig.~\ref{fig2}, $4~\mu s$-long pulses were used, giving a excitation pulse bandwidth of $\Delta \omega_{\textrm{pulse}}/2\pi\approx250$\,kHz. This is slightly smaller than the resonance linewidth $\kappa/2\pi=350$\,kHz, which is thereby barely filtering the input pulses and $\Delta\omega \approx \Delta \omega_{\textrm{pulse}}$. 

ID is caused by the fraction of excited erbium ions which has an effective concentration $\tilde{\rho}=\rho\Delta\omega/\Gamma$.  As $\Gamma$ is of the order of $10$\,MHz, $\tilde{\rho}\sim1.7\times10^{11}$~cm$^{-3}$ and the distance between two excited erbium ions is typically $2~\mu$m. Here, excited spins are in the bulk (few tens of $\mu$m away from the surface) and those within a few $\mu$m distance from each other experience approximately the same Rabi angles. Because spins contributing mostly to the Hahn-echo undergo rotations of first $\theta_1\sim\pi/2$ then $\theta_2\sim\pi$, their excited neighbours also rotate with similar angles and $\sin{(\theta_2/2)}^2$ can be approximated to 1.

For $\varphi=47^{\circ}$, the inhomogeneous linewidth $\Gamma/2\pi$ was measured to be $10$\,MHz (Fig.~\ref{fig1}E) and equation~\ref{T2_ID} gives $T_{2,\text{ID}}\sim400$~ms. This value is more than one order of magnitude larger than the measured coherence time of Fig.~\ref{fig2} and indicates that ID is not the dominant source of decoherence at this angle.

However, the contribution of ID increases when the spin linewidth $\Gamma$ gets narrower or when the excitation bandwidth $\Delta \omega$ increases. To evidence the presence of ID, the coherence time measurement of Fig.~\ref{fig2} can be repeated at $\varphi\sim31^{\circ}$, where the spin linewidth is the narrowest, and also with a different resonator with a broader resonance linewidth. 

\begin{figure}[tbh!]
\centering
\includegraphics[width = 8cm]{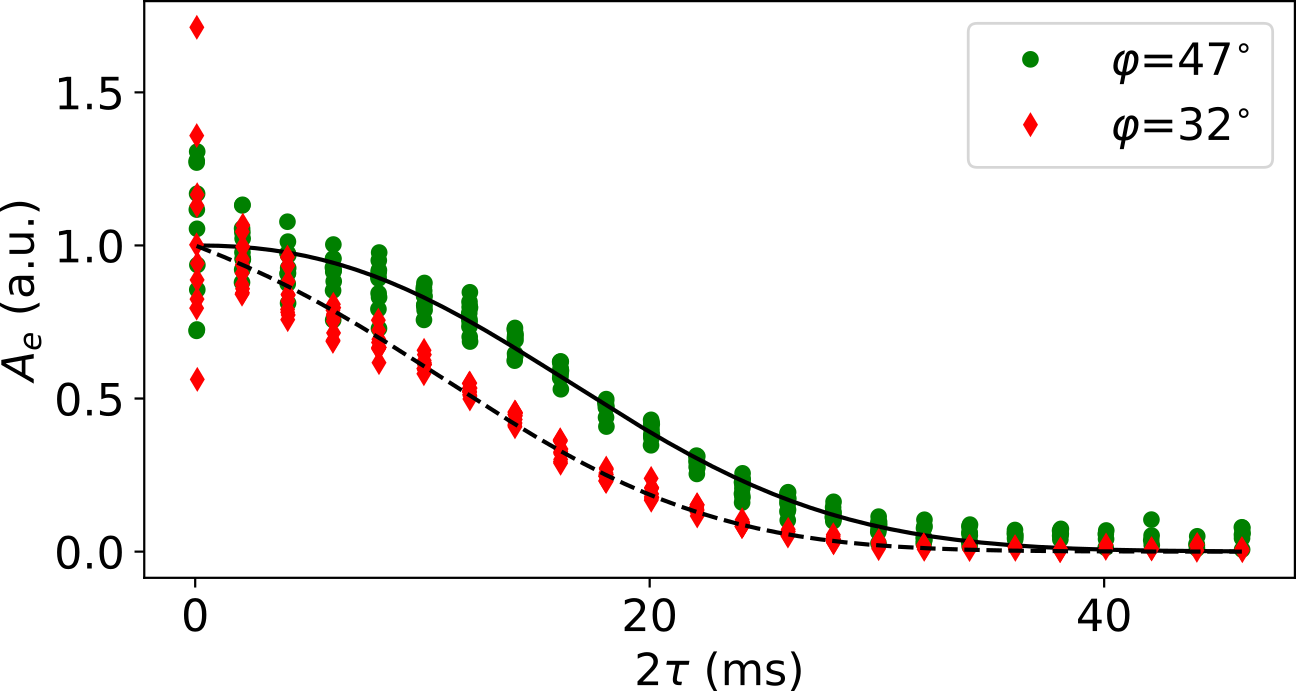}
\caption{\textbf{Electron spin coherence measurements at 10 mK for two magnetic field orientations.} The measurements are performed at $\varphi=47^{\circ}$ (circles) and $\varphi=32^{\circ}$ (diamonds), with a resonance at $\omega_0/2\pi=7.502$\,GHz and with linewidth $\kappa/2\pi=580$\,kHz. Each packet of points is sampled with $\Delta \tau=2~\mu$s in order to account for ESEEM. The data are averaged in magnitude. The solid black line is a fit with $A_e^2=e^{-2[(2\tau/T_{2,n})^{x_n}+(2\tau/T_{2,p})^{x_p}]}+C$ where $T_{2,n}=27.2$\,ms and $x_n=2.74$ are taken from the CCE simulation. The fit yields $T_{2,p}=28\pm1$\,ms and $x_p=2.1\pm0.2$. The dashed black line is a fit which includes ID, $A_e^2=e^{-2[(2\tau/T_{2,n})^{x_n}+(2\tau/T_{2,p})^{x_p}+2\tau/T_{2,\text{ID}}]}+C$. At $\varphi=32^{\circ}$, the CCE simulation predicts $T_{2,n}=24.7$\,ms and $x_n=2.74$. $T_{2,p}$ and $x_p$ are fixed from the fitted values at $\varphi=47^{\circ}$. Fitting the remaining ID contribution yields $T_{2,\text{ID}}=33\pm2$\,ms. The offset $C$ has been subtracted from data and fits (square roots of negative noise on $A_e^2$ are obviously absent from the plot).}
\label{figT2_ID}
\end{figure}

Fig.~\ref{figT2_ID} shows this measurement for a resonance linewidth of $\kappa/2\pi=580$\,kHz with two field orientations $\varphi=47^{\circ}$ and $\varphi=32^{\circ}$. Here, $1~\mu$s-long pulses where used, giving an pulse bandwidth of $\Delta\omega_{\textrm{pulse}}\approx1$\,MHz. The resonance linewidth is $\kappa/2\pi=580$\,kHz hence the excitation bandwidth is limited by $\kappa$: $\Delta \omega \approx \kappa$.

At $\varphi=47^{\circ}$, $\Gamma=11$\,MHz and equation~\ref{T2_ID} yields $T_{2,\text{ID}}\sim190$~ms. At $\varphi=32^{\circ}$, $\Gamma=1.8$\,MHz and equation~\ref{T2_ID} yields $T_{2,\text{ID}}\sim31$~ms.

The resonance linewidth of this measurement is broader than the one used to measure the data of Fig.~\ref{fig2} and the pulses are shorter. As a consequence, high frequencies of ESEEM are less filtered and ESEEM is stronger. Therefore the data are taken in packets of points spaced by $\Delta \tau=2~\mu$s in order to sample the ESEEM properly. The data is fitted in magnitude with the three mentioned decoherence mechanisms,
\begin{equation}
    A_e^2(2\tau)=e^{-2[(\frac{2\tau}{T_{2,n}})^{x_{n}}+(\frac{2\tau}{T_{2,p}})^{x_{p}}+\frac{2\tau}{T_{2,\text{ID}}}]}+C,
\end{equation}
where $T_{2,n}$ and $x_{n}$ correspond to spectral diffusion due to the nuclear spin bath, $T_{2,p}$ and $x_{p}$ to spectral diffusion due to other paramagnetic impurities and $T_{2,\text{ID}}$ to instantaneous diffusion.

At $\varphi=47^{\circ}$, ID is negligible as calculated above. The CCE simulation gives $T_{2,n}=27.2$\,ms and $x_{n}=2.74$. Fitting the paramagnetic contribution yields $T_{2,p}=28\pm1$\,ms and $x_{p}=2.1\pm0.2$. This value is smaller than what is shown in Fig.~\ref{fig2}B where $T_{2,p}\sim40$\,ms at $10$\,mK and could be due to the fact that these data were taken in different runs. The data of Fig.~\ref{figT2_ID} yields a total coherence time of $T_2=20.5\pm0.2$\,ms instead of $23.2\pm0.5$\,ms.

At $\varphi=32^{\circ}$, the CCE simulation gives $T_{2,n}=24.7$\,ms and $x_{n}=2.74$. The paramagnetic contribution can be taken as independent of angle $\varphi$ because these impurities have a dilute concentration and are mostly in tetragonal sites with their g-factor being independent of $\varphi$. Hence we keep the values extracted at $\varphi=47^{\circ}$, namely $T_{2,p}=28$\,ms and $x_{p}=2.1$. Fitting only the remaining contribution from ID yields $T_{2,\text{ID}}=33\pm2$\,ms which is close to the estimation above and confirms that ID is not negligible at angles where the erbium inhomogeneous linewidth is the narrowest.

\subsection{Longitudinal relaxation ($\mathbf{T_1}$) measurements}

\begin{figure}[tbh!]
\centering
\includegraphics[width = 8cm]{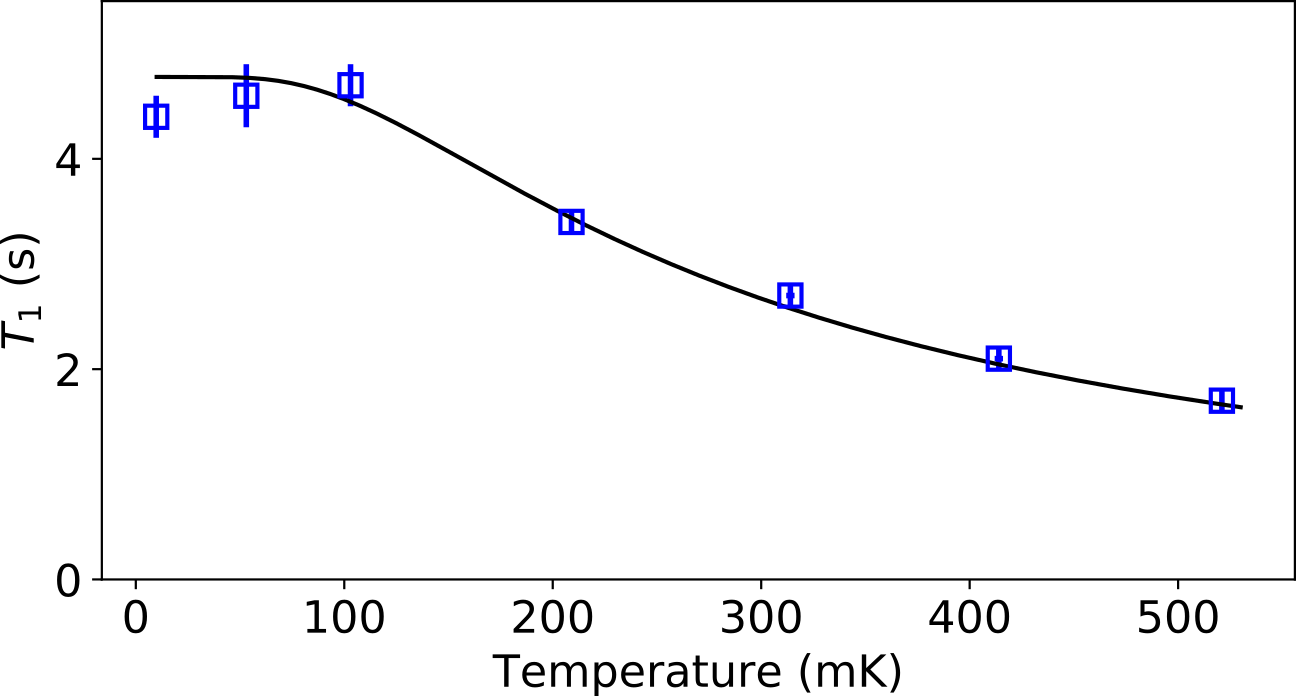}
\caption{\textbf{Longitudinal relaxation time as a function of temperature.} Relaxation time at high input pulse power as a function of cryostat temperature (squares). The solid line is a fit with $T_1=T_{1,\text{0K}}\tanh{(\hbar \omega_0/2k_B T)}$ as is predicted for the spin-lattice relaxation time at low temperature. The fit yields $T_{1,\text{0K}}=4.8\pm0.1$\,s.}
\label{T1_vs_temp}
\end{figure}

We measure the relaxation time $T_1$ as a function of cryostat temperature, with sufficient power to address spins in the bulk of the material (Fig.~\ref{T1_vs_temp}). We observe that the temperature dependence is well-fitted by the direct-phonon process~\cite{abragam_electron_2012}
\begin{equation}
T_1=T_{1,\text{0K}}\tanh{\frac{\hbar\omega_0}{2k_0T}},
\end{equation}
where $T_{1,\text{0K}}$ is the extrapolated relaxation time at zero temperature, confirming that multi-phonon processes are not relevant at sub-Kelvin temperatures. 

\subsection{CPMG enhanced $\mathbf{T_1}$ measurements}

Except for the relaxation measurements presented in Fig.~\ref{fig3}(A,B), all data presented in the manuscript are taken with high pulse powers. For the relaxation measurements in Fig.~\ref{fig3}(A,B), however, the signal-to-noise ratio (SNR) is insufficient for Hahn-echo measurements at low pulse powers. It is therefore necessary to use CPMG sequences to enhance the SNR in this low power regime. Thus the usual Hahn echo sequence ($\pi/2-\tau-\pi-\tau-\textrm{echo}$) is followed by a CPMG sequence consisting in a chosen number $N$ of refocusing pulses ($\tau-\pi-\tau-\textrm{echo}$)$\times N$~\cite{albanese_radiative_2020}. 

In the measurement presented in Fig.~\ref{fig3}(A,B), $\tau$ is chosen to be $30~\mu s$, the spacing between the CPMG pulses is $2\tau=60~\mu s$ and $N$ is set to $332$. The spin-echo amplitude is then computed as the weighted average of the $N+1=333$ echoes. Here the weights are measured from the integral of each refocused echo when no initial inversion pulse is applied. 

The data presented in Fig.~\ref{fig3}C is recorded without CPMG pulses, as the high pulse powers give sufficient SNR with just single-shot Hahn-echoes.

\subsection{Simulations of spin-relaxation power dependence}

The numerical simulations presented in Fig.~\ref{fig3}B take into account a distribution of Larmor frequencies $\delta$ with respect to the resonator frequency and a distribution of spin-resonator coupling constants $g_0$, the two distributions being independent~\cite{ranjan_pulsed_2020}.
\begin{itemize}
    \item for the $2~\mu m$ wide inductance wire, there are 420 discrete frequency bins taken with uniform spacing between $-4\kappa$ and $4\kappa$, where $\kappa/2\pi=185$\,kHz, and $120$ values of coupling strength $g_0/2\pi$, equally spaced between $1$ and $1000$\,Hz,
    \item for the $5~\mu m$ wide inductance wire, there are 480 discrete frequency bins taken with uniform spacing between $-3.5\kappa$ and $3.5\kappa$, where $\kappa/2\pi=350$\,kHz, and $120$ values of coupling strength $g_0/2\pi$, equally spaced between $0.5$ and $500$\,Hz.
\end{itemize}

Each spin with frequency detuning $\delta$ and coupling constant $g_0$ relaxes with rate $\Gamma=\Gamma_P+\Gamma_{sl}$, where $\Gamma_{sl}$ is the spin-lattice relaxation rate and $\Gamma_P=\frac{\kappa g_0^2}{\kappa^2/4+\delta^2}$ is the Purcell relaxation rate~\cite{bienfait_controlling_2016}. For spins at a distance greater than approximately $15~\mu$m from the inductance wire, the relaxation is dominated by $\Gamma_{sl}$, whereas for spins located closer to the indutance wire, it is dominated by $\Gamma_P$.

\begin{figure}[tbh!]
\centering
\includegraphics[width = 8cm]{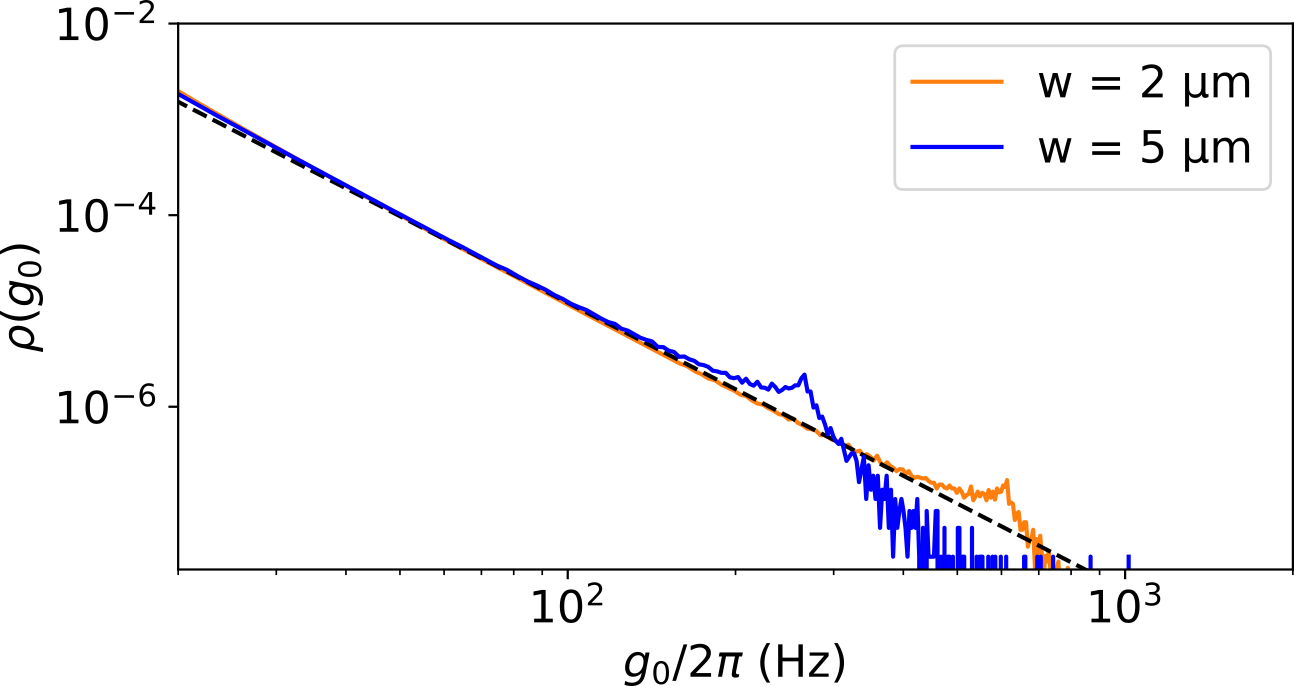}
\caption{\textbf{Spin-resonator coupling distribution}. The coupling distribution is shown in logarithmic scale for the $2~\mu m$ and $5~\mu m$ wide inductor resonators simulated at $\varphi=30^{\circ}$. The COMSOL simulation takes into account spins located in a surface $L_y\times L_z=400\times200~\mu$m$^2$ below the inductance wire. Dashed black line is a fit with $\rho(g_0)\propto g_0^{-3}$. The peak at high coupling corresponds to spins located close to the wire.}
\label{rho_g0}
\end{figure}

The inhomogeneous absorption linewidth at $\varphi=30^{\circ}$ is $\Gamma/2\pi=2$\,MHz, which is nearly one order of magnitude wider than the broadest resonance linewidth. Thus the spin frequency distribution $\rho_\delta$ is taken as constant. The coupling constant distribution $\rho_{g_0}$ behaves approximately as $1/g_0^3$ at low $g_0$ and shows a peak at high $g_0$ due to the spins located close to the wire. The exact profile is calculated using a COMSOL simulation of the magnetic field $\mathbf{B_1}$ generated around the inductance wire with a $1$\,A current. The simulation result must then be rescaled by $\delta I=\omega_0\sqrt{\frac{\hbar}{2Z_0}}$, the rms vacuum fluctuation of the current in the resonator. The resonator impedance $Z_0\sim40~\Omega$ is simulated using the software Ansys HFSS. The computation of the coupling $g_0$ is shown in Fig.~\ref{fig1}C at $\varphi=51^{\circ}$ (i.e. when the magnetic field $\mathbf{B_0}$ is applied parallel to the wire) and the $g_0$ histogram at $\varphi=30^{\circ}$ used for the simulations is shown in Fig.~\ref{rho_g0}.

For simplicity, the simulated pulse sequence for obtaining $T_1$ is the inversion recovery sequence ($\beta-T-\beta/2-\tau-\beta-\tau-\textrm{echo}$), even when CPMG sequences have been used to measure it.

\newpage
\section{Theoretical part}

\subsection{Model}

The spin Hamiltonian can be written
\begin{equation}
H=H_{\mathrm{Er}}+H_{\mathrm{n}}+H_{\mathrm{int}}.  \label{H}
\end{equation}%

The first term,
\begin{equation}
H_{\mathrm{Er}}={\mu _{B}}\mathbf{B}_{0}\cdot \mathbf{g}\cdot \mathbf{S},
\label{H_Er}
\end{equation}
is the Zeeman energy of the effective spin-1/2 ${\mathbf{S}}$ of the zero-nuclear-spin isotopes of Er$^{3+}$ under a magnetic field ${\mathbf{B}}_{0}$, where $\mu _{B}$ is the Bohr magneton and the g-factor tensor $\mathbf{g}$ has a diagonal form in the crystal frame with $\text{g}_{\perp}=\text{g}_{aa}=\text{g}_{bb}$ and $\text{g}_{\parallel }=\text{g}_{cc}$. 

The second term,
\begin{equation}
H_{\mathrm{n}}={\text{g}_{\mathrm{n}}\mu _{\mathrm{n}}}\sum_{i}\mathbf{B}%
_{0}\cdot \mathbf{I}_{i}+\sum_{i<j}\mathbf{I}_{i}\cdot {\mathbb{D}}%
_{ij}\cdot \mathbf{I}_{j},  \label{H_bath}
\end{equation}
is the energy of the bath of $^{183}$\textrm{W} nuclear spins ($I_{i}=1/2$), where $\text{g}_{\mathrm{n}}$ is the g-factor of $^{183}$W nuclear spins, $\mu _{\mathrm{n}}$ is the nuclear magneton, $\mu$ is the vaccum permeability, and ${\mathbb{D}}_{ij}=\mu/(4\pi)\text{g}_{\mathrm{n}}^{2}\mu_{\mathrm{n}}^{2}\left( r_{ij}^{-3}-3\mathbf{r}_{ij}\mathbf{r}_{ij}/r_{ij}^{5}\right) $ with ${\mathbf{r}}_{ij}={\mathbf{r}}_{j}-{\mathbf{r}}_{i}$ the displacement between the $i$-th and $j$-th nuclear spins. 

The third term,
\begin{equation}
H_{\mathrm{int}}=\sum_{i}\mathbf{S}\cdot {\mathbb{A}}_{i}\cdot \mathbf{I}_{i},
\label{H_int}
\end{equation}
is the dipolar hyperfine interaction, where ${\mathbb{A}}_{i}=\mu/(4\pi)\mu _{B}\text{g}_{\mathrm{n}}\mu _{\mathrm{n}}\left[\mathbf{g}/r_{i}^{3}-3\left( \mathbf{g}\cdot \mathbf{r}_{i}\right) \mathbf{r}_{i}/r_{i}^{5}\right] $.

We set the $z$-axis along ${\mathbf{B}}_0$ (in the $ab$-plane with an angle $\varphi$ from the $a$-axis). Since the electron Zeeman energy is much stronger than the hyperfine interaction and the nuclear Zeeman energy is much stronger than the nuclear dipolar interaction, we make the secular approximation, dropping the terms that do not conserve the Zeeman energies, such as the $S_{x/y}$ terms, the $I_{x/y}$ terms, which induce electron spin echo enveloppe modulation (ESEEM), and the $I_i^{x/y}I_j^z$ terms. 

Therefore, the Hamiltonian becomes a pure dephasing model
\begin{equation}
H\approx|1\rangle\langle 1|\otimes H^{(+)}+|0\rangle\langle 0|\otimes
H^{(-)},
\end{equation}
with the central-spin-conditional bath Hamiltonian%
\begin{equation}
H^{\left( \pm \right) }=\pm\frac{ \omega}{2} \pm\frac{1}{2} \sum_{i}%
\hat{\mathbf{z}}\cdot{\mathbb{A}}_i\cdot\hat{\mathbf{z}} I_{i}^{z}+H_{%
\mathrm{n}}.  \label{H_+-}
\end{equation}

The Hahn echo signal is
\begin{equation}
\begin{gathered}
A_e\propto \mathcal{L}\left( 2\tau \right) =\\
\mathrm{Tr}\left( \rho _{\mathrm{n}%
}e^{iH^{\left( -\right) }\tau }e^{iH^{\left( +\right) }\tau }e^{-iH^{\left(
-\right) }\tau }e^{-iH^{\left( +\right) }\tau }\right) ,
\end{gathered}
\end{equation}
where $\rho _{\mathrm{n}}$ is the initial bath density matrix (which we choose as the infinite high-temperature thermalized state since the nuclear Zeeman energy is much less than 10~mK).

\subsection{Cluster correlation expansion}
The central spin coherence is calculated using the cluster correlation expansion (CCE)~\cite{yang_quantum_2008}, in which the decoherence caused by a cluster of $M$ bath spins $(1,2,\ldots ,M)$ is denoted as ${\mathcal{L}}_{1,2,\ldots ,M}$. The irreducible correlation of a cluster is defined recursively as $\tilde{\mathcal{L}}_{j}={\mathcal{L}}_{j}$, $\tilde{\mathcal{L}}_{i,j}\equiv {{\mathcal{L}}_{i,j}}{\tilde{\mathcal{L}}_{i}^{-1}\tilde{\mathcal{L}}_{j}^{-1}}$, etc., that is, the decoherence function divided by all irreducible correlations of all sub-clusters. For the $M$-order truncation (CCE-$M$), the calculation takes into account the irreducible correlations up to the clusters of $M$ spins, ${\mathcal{L}}\approx {\mathcal{L}}^{(M)}$, with
\begin{equation}
\begin{gathered}
{\mathcal{L}}^{(M)}= \\
\prod_{i_{1}}\tilde{\mathcal{L}}%
_{i_{1}}\prod_{j_{1}<j_{2}}\tilde{\mathcal{L}}_{j_{1},j_{2}}\cdots
\prod_{k_{1}<k_{2}\cdots <k_{M}}\tilde{\mathcal{L}}_{k_{1},k_2,\ldots
,k_{M}}.  
\end{gathered}
\end{equation}
With the secular approximation, the CCE-1 contribution (decoherence due to single-spin dynamics, which also causes the ESEEM for relatively strongly coupled nuclear spins) vanishes.

In the simulation, we place the $^{183}$W nuclear spins (with a natural abundance $p_{\mathrm{n}}=0.145$) randomly on the CaWO$_{\text{4}}$ lattice sites of tungsten ions and the Er$^{3+}$ ions randomly substituting Ca$^{2+}$. The bath includes all nuclear spins within a sphere of radius of $11$~nm around the central spin. We numerically checked that a larger bath size produces nearly the same result. The numerical simulation also show that the simulation using one specific spatial configuration of $^{183}$W nuclear spins in the lattice is nearly the same as ensemble average over many ($50$) different spatial configurations.
We have checked the convergence of the CCE and found that CCE-3 and CCE-2 produce nearly identical results.

\subsection{Simulation results}

\begin{figure}[tbh!]
\centering
\includegraphics[width=8cm]{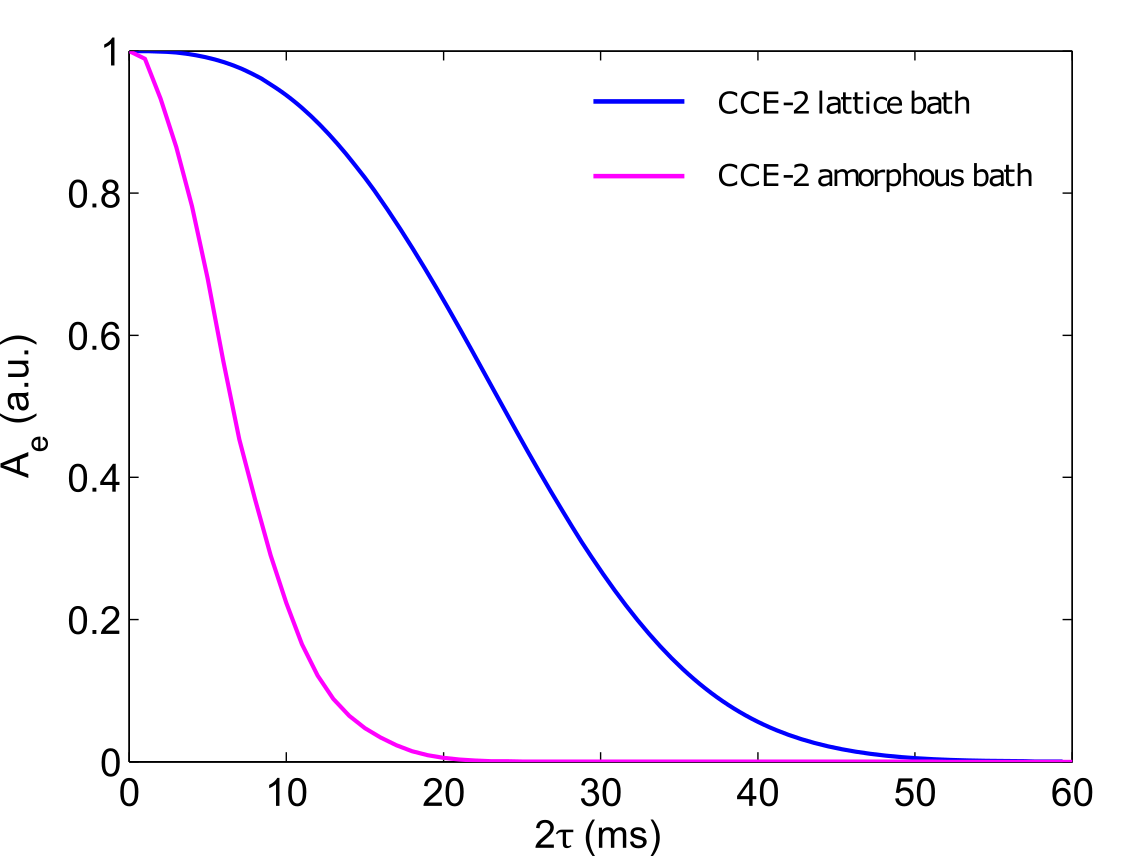}
\caption{\textbf{CCE-2 simulation of the spin decoherence.} 
The results are fitted by $A_{e}=e^{-(2\tau/T_2)^x}$, 
with $T_{2}=27$~ms and $x=2.7$ for the decoherence in a lattice spin bath (blue line) and $T_{2}=8.05$~ms and $x=1.88$ for the decoherence in an amorphous bath (pink line). The parameters of the simulations are the magnetic field $B_0=67$~mT, the field orientation $\varphi=46.5^{\circ}$ and the temperature $T= 10$~mK.}
\label{fig:comparison}
\end{figure}

Fig.~\ref{fig:comparison} compares the CCE-2 simulation in a lattice spin bath according to CaWO$_4$ crystal structure and in an amorphous bath, where the $^{183}$W have same concentration but are placed randomly in space, in order to compare with~\cite{kanai_generalized_2021}. In the amorphous case, the decoherence is significantly faster than in a lattice bath. Such difference can be understood from the fact that the lattice structure sets a lower bound on the distance between nuclear spins, which has a sizeable  effect when the spin concentration is not too small.

\end{document}